\documentclass[twocolumn]{aastex701}
\usepackage{amsmath}

\begin{document}

\title{The \texttt{ViSta} method for stacking in the Fourier domain\\ and its application to the dusty star-forming galaxies in the ALMA Science Archive}

\author[0000-0001-9301-5209]{Martina Torsello}
\affiliation{Scuola Internazionale Superiore di Studi Avanzati, Via Bonomea 265, 34136 Trieste, Italy}
\email[show]{mtorsell@sissa.it}

\author[0000-0002-0375-8330]{Marcella Massardi}
\affiliation{INAF - Istituto di Radioastronomia - Italian ALMA Regional Centre,
Via Gobetti 101, 40129 Bologna, Italy}
\affiliation{Scuola Internazionale Superiore di Studi Avanzati, Via Bonomea 265, 34136 Trieste, Italy}
\email{massardi@ira.inaf.it}

\author[0000-0003-0995-5201]{Elisabetta Liuzzo}
\affiliation{INAF - Istituto di Radioastronomia - Italian ALMA Regional Centre,
Via Gobetti 101, 40129 Bologna, Italy}
\email{e.liuzzo@ira.inaf.it}  

\author[0000-0002-7472-7697]{Gayathri Gururajan}
\affiliation{Scuola Internazionale Superiore di Studi Avanzati, Via Bonomea 265, 34136 Trieste, Italy}
\affiliation{Institute for Fundamental Physics of the Universe (IFPU), Via Beirut 2, 34014 Trieste}
\email{ggururaj@sissa.it}  

\author[0000-0001-9808-0843]{Francesca Perrotta}
\affiliation{Scuola Internazionale Superiore di Studi Avanzati, Via Bonomea 265, 34136 Trieste, Italy}
\email{perrotta@sissa.it}  

\author[0000-0002-4882-1735]{Andrea Lapi}
\affiliation{Scuola Internazionale Superiore di Studi Avanzati, Via Bonomea 265, 34136 Trieste, Italy}
\affiliation{INAF - Istituto di Radioastronomia, Via Gobetti 101, 40129 Bologna, Italy}
\affiliation{Institute for Fundamental Physics of the Universe (IFPU), Via Beirut 2, 34014 Trieste}
\affiliation{Istituto Nazionale Fisica Nucleare (INFN), Sezione di Trieste, Via Valerio 2, 34127 Trieste, Italy}\email{lapi@sissa.it}  


\begin{abstract}
We present \texttt{ViSta}, a Visibility Stacking method to combine interferometric observations in the Fourier domain at radio to sub-millimeter wavelengths for galaxies.
The goal of our method is to maximize the exploitation of available archival interferometric data. By stacking visibilities of galaxies with secure spectroscopic redshifts directly in the Fourier domain and transforming them into the rest-frame, we can enhance the stacked signal, suppress noise, and improve image reconstruction thanks to an extended coverage of the visibility domain. The \texttt{ViSta} method is highly flexible, allowing stacking of visibilities regardless of the array configuration or spectral setup. It is effective for both targeted sources and spurious detections offset from the phase center, whether unresolved or extended, within the field of view of the telescope. We validated the method using simulated interferometric datasets. For point-like sources, we can reconstruct the true emission with approximately 90\% accuracy, obtaining similar results to classical image-plane stacking. In contrast, for faint and extended sources below the noise level, our method can provide a more accurate estimate of the signal compared to traditional image-based approaches.
Finally, we applied \texttt{ViSta} to a sample of dusty star-forming galaxies (DSFGs) observed with the Atacama Large Millimeter/sub-millimeter Array (ALMA) to detect the CO(3-2) emission line . As for the simulated case, we demonstrated that our tool performs better than image-plane stacking when the signal from individual sources is no longer easily detectable, achieving higher SNR. Finally, we outline potential future applications of this stacking approach.

\end{abstract}


\section{Introduction}
\label{sec:introduction}
The radio to sub-millimetric emission of a galaxy is key to investigating the fundamental stages of galaxy evolution as it traces various components of the interstellar medium (ISM), the star formation, and the feedback mechanisms in a galaxy. At cosmic noon (redshift, $z\sim2$), we see the peak of the cosmic star-formation rate density (SFRD), nuclear activity, and the molecular gas content of galaxies. Among the most remarkable objects observed at high redshift at these wavelengths are the dusty star-forming galaxies (DSFGs). These massive (\(M_\star \gtrsim 10^{10}\,M_\odot\)), dust-rich, and highly star-forming systems host extreme  star-formation rates (SFRs) in excess of \(500\,M_\odot\,\mathrm{yr}^{-1}\), contributing approximately \(\sim20\%\) to the total SFRD and \(\sim30\!-\!50\%\) to the stellar mass density of the Universe at \(z \sim 2\!-\!4\) \citep{Casey2014, Madau2014,Talia2021,Behiri2023,Bosi2025}. 
In addition to their continuum emission at FIR-radio wavelengths, characterized by a combination of synchrotron radiation (associated with nuclear activity and stellar processes) and thermal dust emission (heated by young stars in star-forming regions), DSFGs present rich chemistry associated with gaseous and molecular phases of the ISM. Spectral lines such as CO and [C\,\textsc{ii}] trace different phases of the ISM in DSFGs, providing insights into molecular gas reservoirs, kinematics, and excitation conditions. 

However, even for intrinsically bright DSFGs, the signal in individual spectral channels can be very weak (often of the order of only a few tens of $\mu$Jy), making the detection of line emission challenging at high redshift. To overcome this limitation, strong gravitational lensing is sometimes exploited to increase both the observed flux and the apparent angular resolution of these galaxies \citep{Negrello2010,giulietti2024, gururajan22, Vieria2013}, enabling the study of intrinsically faint dusty star-forming galaxies that would otherwise be inaccessible. In addition, spatial and/or spectral stacking techniques are employed to boost the signal-to-noise ratio (SNR). Image-domain stacking has been broadly and successfully used at these wavelengths to recover higher line flux and retrieve the average properties of the ISM of selected samples of high$-z$ extragalactic sources, which would otherwise remain inaccessible in low-sensitivity images \citep{knudsen2005, carilli2008, pannella2009, karim2011, delhaize2013, decarli2014, spilker2014, lindroos2015, jolly2020, vernstrom2021, neumann2023}.

Even for a sensitive telescope such as the Atacama Large Millimeter Array (ALMA), the observation of most of the spectral emission from fainter targets remains extremely time-consuming, and frequently, only the brightest features (typically CO transitions between \(J = 3\) and \(J = 6\), [C\,\textsc{ii}], and the dust continuum) can be detected. As a consequence, other potentially important spectral lines are often not searched for or remain undetected in observations.  

For instance, HCN and HCO\(^+\) are effective tracers of dense molecular gas, as they exhibit strong correlations with star-formation rates across various populations of galaxies \citep{imanishi23}. H$_2$O lines serve as excellent tracers of warm, dense gas, offering insight into star-forming regions \citep{perrotta23}. SiO and CS are commonly used as tracers of shocks in the interstellar medium, as their enhanced emission is typically associated with shocked gas produced by processes such as outflows, cloud–cloud collisions, or galaxy interactions \citep{Imanishi2018, Martin2015}. Methanol (CH$_3$OH) can be released from icy grain mantles under similar conditions, indicating dynamic processes such as outflows or cloud-cloud collisions \citep{huang24}.
 
Stacking of archival datasets could be a gold mine for investigating the DSFG population and improving the detectability of the spectral lines discussed above. It is fundamental to note that archival resources are often inhomogeneous, as a result of many projects with different scientific motivations and varying telescope settings (i.e., sensitivity, spatial and spectral resolution); therefore, a standard approach to stacking has usually been to align the images \citep[produced within the same project or even the same mosaic image][]{decarli2014,spilker2014} before stacking.

Radio astronomical images obtained with an interferometer are the result of an inverse Fourier transform of the observed visibility distribution, namely the map of correlated signals from each antenna couple projected over a plane (usually referred to as `\emph{uv}’ or `Fourier domain’) centered on the target position (tagged as `phase center’) \citep[e.g.,]{thomson2017} . The sampling of the \emph{uv}-domain is inherently incomplete due to the finite number of antennas and limited observing time, leading to gaps in the Fourier space. In addition, converting visibility data into an image involves several steps: resampling the \emph{uv}-plane, applying convolution, and performing an inverse Fourier transform, all while accounting for instrumental effects. The resulting image is then used to predict visibilities, which are subtracted from the observed data to generate residuals. Deconvolution is performed to remove the blurring caused by incomplete \emph{uv}-plane sampling and to refine the image by distinguishing true sky features from distortions. This iterative process gradually improves the sky model, addressing incomplete sampling, and enhancing the image reconstruction.

This technique allows for different possibilities in the weighting scheme, in the deconvolution process, and in the pixelization scheme of the resulting image. Therefore, multiple images can be generated from the same dataset, with different approximations and features appearing differently in the various cases.
Furthermore, the noise no longer remains Gaussian upon imaging the visibilities and it corresponds to the telescope response function due to the deconvolution process involved in the imaging. Therefore, uncertainties in the stacking process should be carefully considered.

To overcome these issues, stacking the interferometric data before imaging (i.e., matching the visibilities in the \emph{uv}-domain) is demonstrated to be more efficient than stacking in the image domain (see \citealt{hancock2011, lindroos2015,Knudsen2015,Hill2024}). However, such methods have been restricted to local galaxies or rely on outdated techniques. To effectively perform visibility plane stacking on sources with different redshift ranges and at high redshifts, it is essential to align the data in the rest frame by applying appropriate frequency scaling and redshift corrections. This process must account for several factors: the effective increase of bandwidth in the observed frequency range, the influence of luminosity distance on the observed flux densities, and the required k-corrections to ensure a consistent visibility-domain stacking across data from different redshifts.
To achieve this, we developed the Visibility Stacking tool \texttt{ViSta}, a new method to investigate sources at high-z by exploiting the availability of radio and sub-mm archival interferometric data observed using differing telescope settings and for sources at different redshifts, by combining the data in the visibility domain. 
Thanks to this approach, we can manipulate interferometric datasets directly in the visibility plane, shifting them to the rest-frame while preserving the original file format. This maintains compatibility with the standard calibration and imaging routines of each telescope’s data processing software, ensuring seamless integration of the stacked visibilities into existing analysis pipelines.

The method is presented in \S \ref{sec:method}, and the details of its application to the data collected in the ALMA Science Archive (ASA) are discussed in \S\ref{sec:VisStainASA}.  Processing strategies for different source morphologies are described in \S\ref{sec:morphology}. To verify the functionality and the efficiency of our method in retrieving spectral features from ALMA observations, we have generated simulated data and compared the \texttt{ViSta} results with a more traditional stacking in the image domain (see \S\ref{sec:mock_data}). Applications to selected real ALMA observations to address specific science cases will be detailed in future papers from our collaboration as they go beyond the scope of this paper. However, as an example of application to real data, we report in \S \ref{sec:real_data} the results of stacking to investigate the CO(3-2) line in a small sample of DSFG observations stored in the ASA. Finally, in \S \ref{sec:conclusions} we summarize our approach and findings and discuss its applications.

Throughout this paper, we adopt a flat $\Lambda$CDM cosmology \citep[]{planck20} with round parameter values $h\approx 0.67$, $\Omega_m\approx 0.3$ and $\Omega_{\Lambda}\approx 0.7$.
\section{The \texttt{ViSta} method}\label{sec:method}

The basic principle of stacking is to align and combine multiple observations of faint or undetected sources, such as galaxies, to enhance their collective signal.
As we are interested in spectral features, we match sources for which we know the position and spectroscopic redshift with high confidence. Indeed, during data combination, even small errors of a few percent can cause a flux loss of nearly 50\% and lead to an increasingly significant broadening of the emission line \citep{jolly2020}. 
Successful stacking requires verifying that each dataset’s spatial filtering and phasing are compatible, ensuring the emission is sampled consistently across all observations. In a radio or sub-mm interferometer, the coverage of the \emph{uv} plane for a given antenna configuration depends on the observing wavelength $\lambda$, total observation time, and the range of baselines, from the shortest distance between any pair of antennas $b_{\rm{min}}$ to the longest $b_{\rm{max}}$. As the Earth rotates, the projected baselines between antenna pairs trace different points in the \emph{uv} plane, thereby improving the sampling over time.

Therefore, the array is able to recover emissions from the source components with angular sizes between the observed resolution $\Theta_{\rm{min}}\sim \lambda/b_{\rm{max}}$ and the maximum recoverable scale $\Theta_{\rm{max}}\sim \lambda/b_{\rm{min}}$: all emission from regions with size $\Theta>\Theta_{\rm_{max}}$ is filtered out, while those $\Theta<\Theta_{\rm_{min}}$ are smoothed to the resolution scale. Different array configurations, spectral set-up and observation times sample different angular scales that should be properly matched to reconstruct average properties. 

If a source is observed in a configuration with its angular scale significantly smaller than $\Theta_{\rm{min}}$, it could be approximated to a Dirac $\delta$ in the phase center: its Fourier transform is a constant and therefore the visibility measured amplitude value will be the same over all baselines (i.e. whatever is the array configuration).  

We can also include resolved sources with angular size $\Theta_{\rm s}$ such that $\Theta_{\rm{min}} < \Theta_s < \Theta_{\rm{max}}$, provided that the \emph{uv} plane is uniformly sampled across the dataset. Moreover, any differences in angular scale sampling across the dataset (e.g., due to different array configurations or observation durations) must be carefully characterized. One strategy to mitigate this is to apply tapering during imaging to match the effective resolution and sensitivity to extended emission.  Alternatively, if we want to be even more strict, we can apply tapering that downweights baselines longer than $b > \lambda / \Theta_s$, effectively smoothing the observations so that the source appears point-like at the tapered resolution.

The presence of other signals in the same field of view (FoV) might contaminate the visibilities, thus in the case of bright contaminating sources, the measured signal is no longer consistent with a constant. Similar cases must be discarded to avoid undesired contamination.

In addition, we aim to exclude high-resolution observations of strongly lensed sources, as well as sources that are significantly more resolved than the rest of the sample. The possible approaches we may adopt for handling these cases, which are still under investigation, are described in the subsequent sections. Therefore, we select only unlensed and moderately resolved sources that meet the above criteria.
Once we have selected a well‐defined, homogeneous sample of sources, we can perform stacking directly in the visibility domain. To combine the visibilities, before performing the Fourier transform, it is essential that the \emph{uv}‐plane sampling remains consistent across all sources, ensuring a uniform angular reference scale. Since each visibility intrinsically encodes its observing wavelength, coherence can only be achieved by reprojecting every visibility to a single, common reference, i.e., the rest‐frame wavelength  $\lambda_{\rm{RF}}$. Spectral matching is performed by shifting each dataset to the rest-frame wavelength $\lambda_{\rm{RF}}=\lambda/(1+z)$. Therefore, the input dataset at wavelength $\lambda$ is equivalent to being observed at $\lambda_{\rm{RF}}$ by the same array but with a more compact configuration with antennas at distances $b/(1+z)$. Furthermore, the spectral resolution is also increased to $\Delta\lambda_{\rm{RF}}=\Delta\lambda/(1+z)$. Once coherence has been established across all sources, their visibilities can be concatenated to achieve consistent coverage of the same \emph{uv}-plane.

The core of the data pre-processing in the \texttt{ViSta} method is to adjust the telescope settings and manipulate the interferometric data to simulate an observation of the same identical source as if it were at redshift $z=0$, but emitting at its rest-frame wavelength with a rest-frame luminosity density $L_{\rm{RF}}=S_{\rm obs} 4\pi D_{\rm{L}}^2/(1+z)$, where $S_{\rm obs}$ is the observed flux density and $D_{\rm{L}}$ the luminosity distance.
Once every dataset to be stacked has been shifted to the rest frame and properly centered, we also apply a correction for the primary beam attenuation that the source would have experienced at its original position in the field. This ensures that the flux density is accurately recovered after the spatial shift to the phase center. Visibilities must then be spectrally regridded to a finer common spectral grid. We adopt the spectral sampling of the dataset with the highest native resolution in order to prevent undersampling and preserve the fidelity of narrow spectral features.

Once the pre-processing steps (centering with primary beam correction, rest-framing, and spectral alignment) are complete, we can concatenate all datasets and treat them as if they were multiple epochs or configurations observing the same target. Note that in this process, there is no loss of information about the spectral profile and each dataset is intrinsically weighted for its thermal noise limits.  

After concatenation, the resulting dataset is fully processed and ready for scientific analysis. Additional post-processing steps can be applied depending on the specific scientific goals and the data’s characteristics. It is important to note that these steps are optional, as the core stacking procedure formally concludes with the concatenation, while any further operations are left to the user’s discretion.

For example, once the stacking is complete, we can perform continuum subtraction to isolate the spectral line signals and simultaneously estimate the average continuum flux of the sample. In this way, we also estimate the average continuum flux of the sample. However, continuum removal may not be recommended when the line emission does not significantly stand out from the continuum, as this could lead to unintended suppression of the signal.

Similarly, the weighting step is another optional operation used to correct for uneven spectral coverage. To avoid statistical biases, which can arise when spectral windows do not perfectly overlap and some channels are averaged more heavily than others, each source can be weighted in each channel to account for possible repetitions in the concatenation, and the resulting visibilities are re-normalized channel by channel to achieve a more uniform combination of the datasets. This weighting may not be advantageous in cases of marginal line detections, where it could inadvertently suppress the signal, particularly if nearby lines fall within the stacking window.

Among the post-processing operations, we also include two methods for extracting the final flux. We can measure the spectrum directly by averaging the visibility values (over the whole \emph{uv}-domain) in each channel: the signal would be amplified and the Gaussian noise would be minimized. It is also possible to generate an image out of the concatenated visibilities and measure the luminosity of the stacked source. The resulting image will benefit from an improved \emph{uv}-coverage compared to those obtained from any individual input dataset. While these two approaches are provided as practical tools to recover the bulk of the emitted flux from the stacked signal, the user is free to adopt any alternative method that best suits their scientific objectives.

\section{Applications to the ALMA Science Archive}\label{sec:VisStainASA}

\begin{figure}[!htpb]
\centering
    \includegraphics[trim={4.2cm 0cm 4.2cm 0cm},clip,width=0.403\linewidth]{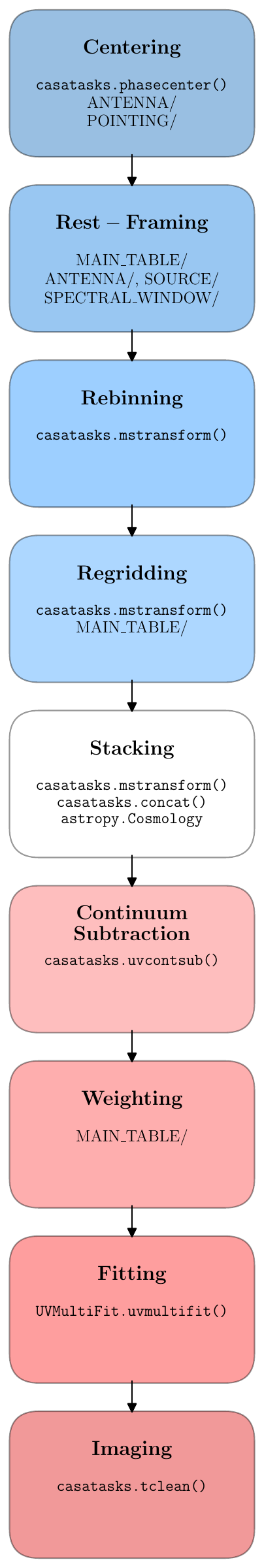}%
    \caption{The \texttt{ViSta} method general workflow. We also provide specifics for ALMA Measurement Sets, listing the functions from CASA or other astronomical libraries and the tables that have been manually modified. The blue color cells refers to the input data editing phase, while the red color cells represents the output extraction steps.}
    \label{flowchart}
\end{figure}

In this section we describe how each step of the \texttt{ViSta} procedure is implemented to deal with ALMA data stored in the ASA. While the described method could, in principle, be applied to any sample of sources with known position, redshift and size, we mainly focus on DSFGs. We selected a sample of $\sim 200$ DSFGs with the criteria that the galaxy falls within the redshift range $z=2$ -- 4.5 and has been observed in at least one ALMA band. Subsamples can be defined depending on specific scientific goals (which we will describe in Sect. \S \ref{sec:conclusions} for our sample of galaxies).

After calibrating each dataset\footnote{We mostly retrieved calibrated datasets by exploiting the EU ARC CalMS service \url{https://almascience.eso.org/tools/eu-arc-network/the-european-arc-calms-service} both for archival products and for the ARI-L project products \citep[]{massardi21}}, visibilities appear in the Measurement Set \citep[MS]{dijkema2019} format: a collection of inter-dependent tables, containing information on the visibilities, telescope settings, applied calibration gains, metadata summarizing the dataset history, among other relevant parameters.
To analyze the ALMA Measurement Sets within the ViSta framework, we mainly employed version 6.6 of the Common Astronomy Software Applications (CASA) package, which is the primary data processing software for ALMA \citep{casa}.
It supports a wide range of operations, including calibration, imaging, and data analysis, making it well-suited for handling our datasets. In addition, we employ several Python libraries to facilitate data manipulation and analysis, including \texttt{Astropy} for managing astronomical data structures and units, \texttt{SciPy} and \texttt{NumPy} for numerical computations, and \texttt{Matplotlib} for data visualization. 

Figure \ref{flowchart} presents a comprehensive overview of the ViSta workflow. At each step, we indicate the specific packages involved and the functions used to modify the data, along with the MS tables that we edited via Python scripts. This figure serves as a visual guide to the data processing pipeline for stacking ALMA data described in the following paragraphs. In the case of other interferometric telescopes, the pipeline will remain conceptually the same, although different software tools will be employed and the data will be adjusted accordingly.

\subsection{Centering}
The centering process ensures that all images are aligned to a common reference position, allowing coherent stacking of multiple observations.
First, we need to determine the precise position of the source. This position can be derived either from the continuum emission of the observation itself, for instance, by fitting a Gaussian to the source emission, or by exploiting a different observation of the same source retrieved from the ALMA archive. Once the accurate coordinates of the source are identified, the phase center of the observation must be shifted to align with this reference position.

This phase center shift is performed using the \texttt{casatasks.phaseshift()} function, which is applied to each input MS of the sample, hereafter referred to as \emph{file.ms}, and applies a phase correction based on the new coordinates (the specified ``phasecenter'') and produces a corrected MS \emph{file.ms.center} where the source is now at the center of the field. This approach ensures that the original data remain untouched.

After the phase center has been shifted, we modify the metadata stored in the FIELD, SOURCE, and POINTING tables of the MS. In particular, we simulate that all sources are located at coordinate zero, which is essential for correctly concatenating multiple datasets. This is done by setting to zero the entries corresponding to the phase direction $\textsc{PHASE\textunderscore DIR}$, reference direction \textsc{REFERENCE\textunderscore DIR}, and delay direction \textsc{DELAY\textunderscore DIR} columns in the FIELD folder. Additionally, we set the source direction entries in the POINTING table to zero, ensuring full consistency across the dataset. These steps guarantee that all sources in the sample are treated as if they were observed at the same position, allowing accurate combination and stacking in the following analysis steps.

Once the source has been shifted to the center of the field, we divide each visibility by the correction for primary beam attenuation that it would have experienced at its original position in the field of view. This correction is crucial to recover the true flux density of the source prior to stacking. The primary beam (PB) response of an ALMA 12-meter antenna can be approximated by a normalized symmetric Gaussian function that depends on the observing frequency $\nu$ and the angular offset $\theta$ from the pointing center:

\begin{equation}
\text{PB}(\theta) = \exp\left[ -4 \ln(2) \left( \frac{\theta \, \nu \, d}{1.2 \, c} \right)^2 \right],
\end{equation}

where $d$ is the antenna diameter and $c$ is the speed of light. This ensures that all sources are placed on a consistent flux scale regardless of their original location in the field, enabling accurate stacking across the sample.

\subsection{Restframing}\label{sec:restframe}

The process of rest-framing the visibilities begins by shifting the observed frequencies to the rest frame. The process of rest-framing the visibilities begins by shifting the observed frequencies to the rest frame. This transformation requires modifying the frequencies, along with all dependent parameters (for example, the channel spectral resolution and the total bandwidth) by applying the corrective factor $(1+z)$. For a typical MS from a given ALMA observation, majority of the parameters that depend directly on the frequency can be found in the \textsc{SPECTRAL\textunderscore WINDOW} folder. These parameters include the central frequencies for each channel,\textsc{CHAN\textunderscore FREQUENCY}, the reference frequency for the spectral window, \textsc{REF\textunderscore FREQUENCY}, the spectral width of each channel, \textsc{CHAN\textunderscore WIDTH}, the effective noise bandwidth of each spectral channel, \textsc{EFFECTIVE\textunderscore BW},the effective spectral resolution of each channel, \textsc{RESOLUTION}, and the total bandwidth for this window, \textsc{TOTAL\textunderscore BANDWIDTH}. We create a copy file called (\emph{file.ms.center.rest}) where we apply all these modifications.

Furthermore, the baseline vectors and antenna positions stored in the MS must also be corrected, since they are expressed in units of wavelength $\lambda$ and thus depend directly on the observed frequency. In particular, the visibility coordinates in the main table \textsc{UVW}, the antenna positions in the \textsc{ANTENNA} table \textsc{POSITION}, and the dish diameter \textsc{DISH\textunderscore DIAMETER} must be divided by the corrective factor $1+z$. In addition, the dish diameter must also be rescaled to preserve the angular resolution as the beam size $\theta \approx \lambda / d$, so adjusting $d$ by $(1+z)$ thus preserves the consistency of $\theta$ between the observed and the rest-frame.

\subsection{Rebinning}
Before merging different datasets, it is essential to ensure consistency along the spectral axis across all MS, meaning that all datasets must share the same spectral resolution and channel width so that spectral features are properly aligned and can be reliably combined. Initially, the spectral resolution of each observation is adjusted using the \texttt{casatasks.mstransform()} tool to match the widest spectral resolution among the sample and returns a rebinned measurement set \emph{file.ms.center.rest.rebin}. This step prevents the introduction of artifacts during subsequent data combination.
To improve spectral smoothness, one can first apply a Hanning smoothing filter with \texttt{casatasks.hanningsmooth()}, which performs a weighted running average of each channel with its two neighbors (0.5 for the central channel, 0.25 for each adjacent channel) and flags the first and last channels, producing \emph{file.ms.center.rest.smooth}. This smoothed MS can be further rebinned with \texttt{casatasks.mstransform()}, yielding \emph{file.ms.center.rest.smooth.rebin}. Note that if the correlator has already pre-averaged the data during calibration, \texttt{hanningsmooth} will detect this and the smoothing step for those spectral windows can be skipped.

\subsection{Regridding}
The next step is to standardize all observations so that they have the same number of channels. This is a prerequisite for using the CASA \texttt{casatasks.concat()} function subsequently, which requires consistent spectral window shapes across the datasets to merge them. In the future, however, the plan is to avoid combining the data with this function, allowing for more flexibility in the choice of spectral regions. Therefore, we adopt a number appropriate for the dataset, constrained by the MS with the fewest, and for each MS we select the spectral window that best overlaps with the others and matches our scientific interests. As a result, some regions of the spectrum will be covered by multiple observations, while others will have fewer contributions. We can decide to limit ourselves exclusively to the spectral region where the entire dataset overlaps, although this would discard a substantial amount of data, or, as we will explain in the following steps, carefully weight each channel according to the number of observations that have a detection in that channel.
Another possibility is to add “dummy” channels to avoid being limited by the size of the final spectral window. For example, suppose we want the largest possible spectral window that covers all frequencies observed in our dataset. To define the final frequency range, we select the minimum frequency from the lowest observed range and the maximum frequency from the highest observed range across all datasets. To standardize the number of channels, we identify the observation with the largest number of channels and use it as a reference. The spectral coverage of the other datasets is then expanded by adding flagged channels at the edges of their spectral axes, ensuring that all observations have the same number of channels without exceeding the overall minimum and maximum frequency limits. To add the missing channels, we first created multiple copies of the original MS, to preserve its full table structure, including all spectral windows and metadata. For each copy, we then run \texttt{casatasks.flagdata()}, the all-purpose CASA flagging task that marks data as invalid, thereby turning every visibility into a ``placeholder'' flagged channel. Next, we merge these flagged MS copies back into the original dataset using \texttt{casatasks.concat()}, which preserves the main structure of its spectral features by adding the new channels on the old one. Finally, we apply \texttt{casatasks.mstransform()} to cut the measurement set and extract our final spectral cube. We trim any excess flagged channels outside the desired overall frequency bounds and regrid the data onto the target channelization, producing a uniformly sampled cube ready for stacking called \emph{file.ms.center.rest.rebin.regrid}.

\subsection{Stacking}

Once the data are prepared, we proceed with the actual stacking operation. Before combining the measurement sets, we rescale the visibilities of each observation by multiplying them by a correction factor to adjust the flux densities to a common reference redshift. Specifically, each visibility is multiplied by
\begin{equation}
w = \frac{D_L^2(z_{\text{source}})}{D_L^2(z_{\text{common}})} \times \frac{1 + z_{\text{common}}}{1 + z_{\text{source}}}
\end{equation}
where $z_{\text{source}}$ is the source redshift, $z_{\text{common}}$ is the reference redshift chosen for the stacking analysis, and $D_L(z)$ is the luminosity distance at redshift $z$. This correction accounts for the difference in luminosity distance and redshift effects, ensuring that flux densities are consistent and comparable across the sample. This rescaling step is optional and can be skipped or adjusted according to the specific scientific goals of the analysis.
After this rescaling, the measurement sets are concatenated using the CASA \texttt{concat()} function to produce a single combined dataset, \emph{concat.ms}.

Furthermore, if a source is gravitationally lensed and remains unresolved or partially resolved in our data, its flux is corrected by dividing the corresponding visibilities by the lensing magnification factor before stacking. This removes the artificial flux boost due to lensing, allowing the stacked dataset to reflect the intrinsic (unlensed) brightness of all sources.

If the data have been properly pre-processed and aligned, the \texttt{concat()} function treats all the observed sources as a single field centered at zero with a unified spectral window. Although the sample is homogeneous, each source is added to \emph{concat.ms} as a separate observation. This structure can complicate operations such as time averaging using \texttt{mstransform}, which cannot average across scans belonging to different observations. Time averaging is also necessary to reduce computational load and to visualize the stacked compact spectra in the \emph{uv}-plane using plotting tools like \texttt{casaplotms}.

To address these issues, we modify the \textsc{OBSERVATION} table so that all scans appear as part of a single observation. First, we determine the global time span of the dataset by extracting the minimum and maximum values from the \textsc{TIME\textunderscore RANGE} column. We then update the first row (observation ID 0) of the \textsc{OBSERVATION} table to cover this full time range. Subsequently, we overwrite the \textsc{OBSERVATION\textunderscore ID} column in the main table, setting all entries to 0. This effectively unifies the dataset under a single observation ID. At this point, we can safely use \texttt{casatasks.mstransform()} to extract only observation ID 0 and discard the others in a new \emph{concat.ms.unified} file, on which we can perform operations like time averaging. 

As discussed previously, the stacking procedure formally concludes with the concatenation step, and any further processing is optional and depends on the specific scientific goals. 

\subsection{Continuum Subtraction and analysis}\label{sec:continuum}

After stacking the data, we may proceed to separate the continuum emission and isolate the spectral line signals. This is achieved using \texttt{casatasks.uvcontsub}, which performs continuum fitting and subtraction in the \emph{uv}-domain. The task estimates the continuum emission by fitting polynomials to the real and imaginary parts of the visibility data over line-free channels specified by the \texttt{fitspw} parameter. Typically, a zero-order polynomial (i.e., a constant) is used to avoid overfitting and to preserve the line flux. The fitted continuum model is then subtracted from the data, resulting in a new MS \emph{concat.ms.unified.contsub}, which contains only the line emission.
Optionally, if the parameter \texttt{write\textunderscore model = True} is set, the task also writes the fitted model in the \textsc{MODEL} column of the main table of the output MS. 
In fact, this process provides the continuum term concatenated over the whole sample, corresponding to the stacked visibility of the continuum signal. By producing a multi-frequency-synthesis (mfs) image, for example with the \texttt{TCLEAN} task, we obtain the stacked continuum map, ready for any population continuum analysis. 

\subsection{Weighting}

During the imaging process, the data are weighted according to the noise in each of the source datasets (i.e., noisier data, coming from shorter integration times, receive less weight). This inverse-variance weighting is usually optimal for maximizing the overall signal-to-noise ratio and is applied automatically during the cleaning step. Moreover, as we have seen in the regridding step, some channels may be covered by fewer observations than others, and we can account for this in the subsequent processing.
To standardize the frequency grid across all observations while regridding, we add flagged data, which leads to regions that could be averaged with inhomogeneous numbers of sources. Therefore, to ensure statistical consistency across the spectral window, we account for the number of observations that cover each frequency and apply corresponding statistical weights to each channel.
The weight assigned to each channel is defined as:
\begin{equation}
w(\nu) = \sqrt{\frac{N_{\mathrm{coverage}}(\nu)}{N}},
\end{equation}
where $N_{\mathrm{coverage}}(\nu)$ is the number of observations that include the frequency $\nu$, and $N$ is the total number of sources in the sample.

To compute $N_{\mathrm{coverage}}(\nu)$, we first construct a uniform frequency array with the same shape as the unified spectral window. For each observation in the dataset, we determine the indices of the grid corresponding to its original frequency range (from its minimum to maximum frequency) and increment the entries in this range by one, accumulating the counts of sources contributing to each frequency bin. This results in an array that encodes the number of overlapping observations per grid frequency. As the centers of the unprocessed spectral channels may not align exactly with the grid points, we interpolate this coverage array onto the precise channel frequencies to obtain a continuous estimate of $N_{\mathrm{coverage}}(\nu)$.

Moreover, if two observations of the same source overlap in some channels, we must reduce the weight of each observation on those channels (by dividing by the number of times the source appears) to avoid multiple counting. This ensures that, statistically, each source contributes only once in a single channel to the final stacked dataset, regardless of how many times it is observed at that frequency across different datasets.

The choice of weighting is flexible and can be adapted according to the specific scientific goals of the analysis. Users may choose to skip this step entirely or modify individual observations, allowing them to apply the weighting scheme that best suits their needs.

\subsection{Spectral line fitting in the visibility domain}

To fit the stacked data from ALMA observations directly on the visibility plane, we utilize the \texttt{UVMultiFit} package, a versatile tool designed to fit models directly on interferometric visibility data within the CASA environment. This approach bypasses the imaging process, enabling more accurate parameter estimation, in particular for unresolved or partially resolved sources.

In our analysis, we employ the function \texttt{UVMultift.uvmultifit()}, selecting either the \texttt{delta} model (corresponding to a point source) or the \texttt{Gaussian} model, depending on the source morphology.

The fitting procedure is performed on a per-channel basis by setting \texttt{OneFitPerChannel = True}, providing an estimate of the flux density and its associated uncertainty for each spectral channel. Additionally, using the \texttt{write = `residuals'} parameter, the residual visibilities are saved in a column of the measurement set. This enables the assessment of the fit quality through residual analysis.

Finally, we plot the fitted flux densities as a function of frequency, providing the spectrum in the visibility domain. We can model this spectrum by fitting a Gaussian profile, facilitating the characterization of spectral features directly from the visibility data.

\subsection{Imaging}\label{method_imaging}

The final step involves transforming the stacked visibilities into the image plane using the \texttt{casatasks.tclean()} task from the CASA software package, which is a versatile imaging tool that reconstructs images from interferometric data by deconvolving the point spread function (PSF) from the observed visibilities.

To image our stacked spectra, we adopt the following parameter settings:

\begin{description}
    \item[Deconvolution Algorithm]: We use the \emph{Hogbom} algorithm, a classic cleaning method suitable for fields dominated by point sources or compact emission. This algorithm iteratively identifies the brightest pixel in the residual image, subtracts a scaled PSF, and updates the model image accordingly.
    
    \item[Weighting Scheme]: We select \emph{natural} weighting to maximize sensitivity. This weighting scheme gives more weight to densely sampled regions in the \emph{uv}-plane, resulting in lower noise levels at the expense of angular resolution. Since our primary goal is to recover the total flux of the stacked source rather than to resolve its detailed spatial structure, the loss in resolution is acceptable for our scientific goals.
    
    \item[Cell Size]: We define the cell size to be $1/5$ of the major axis of the restoring beam. This choice ensures adequate sampling of the PSF.
\end{description}
First, we generate a dirty image (\emph{niter=0}) to check for possible artifacts introduced by the stacking process and to evaluate whether the source is already detected above the noise. We determine the cleaning threshold for the deconvolution based on the average sensitivity expected for the sample, which is derived from the estimated sensitivity $\sigma_k$ of each individual observation \( k \) as:

\begin{equation}
\sigma^2 = \frac{1}{\sum_{k=1}^{N} 1/\sigma_k^2}.
\end{equation}

The threshold is typically set at a level of $3\sigma$. With these parameters, we perform the deconvolution using \texttt{tclean}, which produces the image cube, along with the residuals, model image, point spread function (PSF), and primary beam. These products can then be visualized with an image viewer such as \texttt{CARTA}, to analyze the spatial and spectral properties of the detected emission.

\section{Processing strategies for different source morphologies}\label{sec:morphology}

The visibility–stacking procedure described in the previous sections is fundamentally independent of the intrinsic angular size or complexity of the target sources, provided all the datasets uniformly sample the \emph{uv}-plane between a common minimum baseline ($\Theta_{\rm min}$) and maximum baseline ($\Theta_{\rm max}$). However, source morphology must be taken into account when combining the ensemble of visibilities. Point–like, partially resolved, and extended (or lensed) sources require different treatments, and combining them together requires specific post-processing considerations. Therefore, in Sects. \ref{sec:unresolved_sources}, \ref{sec:partially_solved} and \ref{sec:lensed}, we outline how to handle different source morphologies and describe the corresponding signal extraction strategies for unresolved sources, partially resolved sources, and resolved or lensed sources.

\subsection{Unresolved Sources}\label{sec:unresolved_sources}

In the simplest case, where the entire signal is contained within the synthesized beam, the source can be treated as unresolved. If we choose to extract the flux after imaging, we can simply take the peak intensity within the synthesized beam. Alternatively, for the analysis in the visibility domain, we use \texttt{uvmultifit} with a Gaussian profile: the size parameters are allowed to vary within physically motivated priors, while the position is fixed at the center. This approach provides a more robust estimate than a delta-function model, particularly in the presence of noise or slight deviations from ideal point-like structure.

\subsection{Partially-Resolved Sources}\label{sec:partially_solved}

In cases where the sample includes partially resolved sources, it is important to evaluate whether their intrinsic angular sizes are smaller or larger than the common restoring beam determined by the combined \emph{uv}-coverage. When the beam is at least three times larger than the intrinsic source size, individual emissions are smoothed in the stacking process, permitting treatment analogous to that of point-like sources within the same framework.
However, if the signal-to-noise ratio is low or the data contain a substantial amount of noise, those noisy fluctuations will also be convolved into the beam and propagated into the final stack. This effectively raises the noise floor of the stacked image and can bias any measurement of faint emission. 

In contrast, when the angular extent of the source approaches or surpasses the beam size, its peak image intensity no longer captures the total flux density. In such cases, the appropriate procedure is to explicitly model the spatial emission profile. If the angular size of the source in the band of interest is known, one could measure the flux density directly using \texttt{casatasks.imstat()} within an aperture tailored to the emission; however, since the intrinsic extent typically varies with frequency and the target line may exhibit velocity dispersion across the band, it is preferable to perform a two-dimensional Gaussian fit with \texttt{casatasks.imfit()}. Alternatively, a visibility-domain Gaussian fit with \texttt{uvmultifit} with a Gaussian model and appropriately broad size priors can be used to estimate the total flux density of the stacked source, accounting for the intrinsic source size variation across frequency.

Another strategy is to enforce a common resolution across the sample directly during imaging. This can be achieved using the \texttt{restoringbeam} parameter in the \texttt{tclean} task in CASA, which forces the final image to adopt a beam size corresponding to the worst (i.e., largest) resolution in the dataset. This ensures uniform angular resolution across all sources, allowing a consistent and unbiased comparison within the stacked sample. Finally, the flux of the stacked source can then be estimated with \texttt{casatasks.imfit()} similar to the previous case.

\subsection{Resolved and High-Resolution Sources}\label{sec:lensed}

When sources are well resolved (typically at angular resolutions of $0.2''$ to $0.1''$) and exhibit clearly extended structures, as in the case of high-resolution observations of lensed galaxies, they require additional care to be included in a stacked sample alongside unresolved or partially resolved sources.
An effective strategy to address this is the application of \emph{uv}-tapering. Tapering is a weighting technique applied during imaging that downweights visibilities at longer baselines, effectively reducing the angular resolution of the final image. By applying a tapering function (typically Gaussian in the \emph{uv}-plane), one can smooth high-resolution data to match the angular scale probed by lower-resolution observations. This ensures that all sources in the sample are treated with a consistent effective beam size, thereby enabling a meaningful combination in the stacking process.
Once the appropriate tapering has been applied to smooth the more highly resolved sources to a resolution compatible with the rest of the sample, these sources can be modeled following the same procedure described for partially resolved sources, with an image-plane modeling or tailored visibility-domain fitting with flexible priors. However, these strategies are yet to be further validated and could be refined by future studies. 

\section{Example on simulated data} \label{sec:mock_data}

\subsection{Building simulated ALMA dataset}

To verify the proper functioning of \texttt{ViSta}, we test the validity and functioning of Vista on a sample of 50 simulated galaxies. Each of the mock galaxies is assigned a random value of redshift between 2-4, where we simulate both the continuum and the line emission.
The emission line is modeled as a Gaussian bump superimposed on the continuum, and it is located at the same rest-frame wavelength for all sources.

Then, we use the task \texttt{simobserve}\footnote{\url{https://casaguides.nrao.edu/index.php/Simulating_Observations_in_CASA_6}} in the Common Application Software for Astronomy (CASA) v.6 to simulate ALMA observation for each source. The task accepts as input several parameters like the sky model, the telescope configuration (thus the angular resolution of the observation), the observation date and the total observation time and returns a simulated MS with the same construct of an actual ALMA observation. 

\texttt{simobserve} allows for adding thermal noise according to realistic models of the atmospheric profile for the ALMA site, tabulated in dedicated libraries \footnote{\url{https://cab.inta-csic.es/users/jrpardo/class_atm.html}}, and described as:

\begin{equation}\label{eq:noise}
S_\nu(\text {noise})=\frac{4 \sqrt{2} k}{\eta_{\rm{a}} \eta_{\rm{c}} \pi d ^{2} \sqrt{\Delta \nu \Delta t}} T_{\rm{sys}}
\end{equation}
where $k$ is the Boltzmann constant, $\eta_a$ and are the antenna and the correlator efficiency, $d$ is the diameter of the antennas, $\Delta \nu$ is the observed bandwidth and $\Delta t$ is the integration time. The system temperature, $T_{\rm{sys}}$, comprises contributions from various components and is defined as:

\begin{equation}
T_{\text{sys}} =\  T_{\text{CMB}} + \eta_s T_{\text{atm}} (e^{\tau_\nu} - 1) + (1 - \eta_s) T_{\text{amb}} e^{\tau_\nu} + T_{\text{RX}} e^{\tau_\nu}
\end{equation}

where $T_{\rm{CMB}}$ is the temperature of the cosmic microwave background, $\eta_s$ indicates the spillover efficiency of the antenna, $T_{\rm{atm}}$ is the atmospheric temperature, $\tau_\nu$ denotes the atmospheric opacity at the observed frequency, $T_{\rm{amb}}$ is the ambient temperature and $T_{\rm{RX}}$ represents the receiver temperature. Together, these factors account for the total thermal noise introduced during the observation process. 

However, the corruption procedure in the \texttt{simobserve} task only calculates a single average $T_{\rm{sys}}$ for all visibilities and integration times, without considering the phase delay for each antenna introduced by the atmosphere. As a result, the simulator generates individual noises with an identical standard deviation and the weights and sigma associated with each individual antenna are identical and set to unity. This not only fails to accurately represent a realistic observation, where each antenna is affected differently by the atmospheric contribution, but also does not allow us to exploit the full potential of stacking in the visibility plane (as it does not include random fluctuations of the noise). 

For these reasons, we choose to generate the MS for each mock source without noise and then corrupt them using custom Python functions. These functions add thermal noise, with the functional description given by Equation (\ref{eq:noise}), by exploiting the same libraries as \texttt{simobserve}, but introducing phase delay and fluctuations. The atmospheric temperature and thermal opacity are individually calculated for each antenna, taking into account factors such as antenna elevation and the air mass above each one. Other parameters, such as humidity, ambient temperature, etc., are set based on the values suggested by the ALMA Sensitivity Calculator \footnote{\url{https://almascience.eso.org/proposing/sensitivity-calculator}}, with random variations of up to 10\% applied to key observational parameters, including atmospheric opacity, ground temperature, and receiver temperature, to ensure each simulated observation is unique.

\begin{figure*}[!htpb]
    \centering
    \includegraphics[trim={0.2cm 0cm 0.5cm 0.5cm},clip,width=0.24\linewidth]{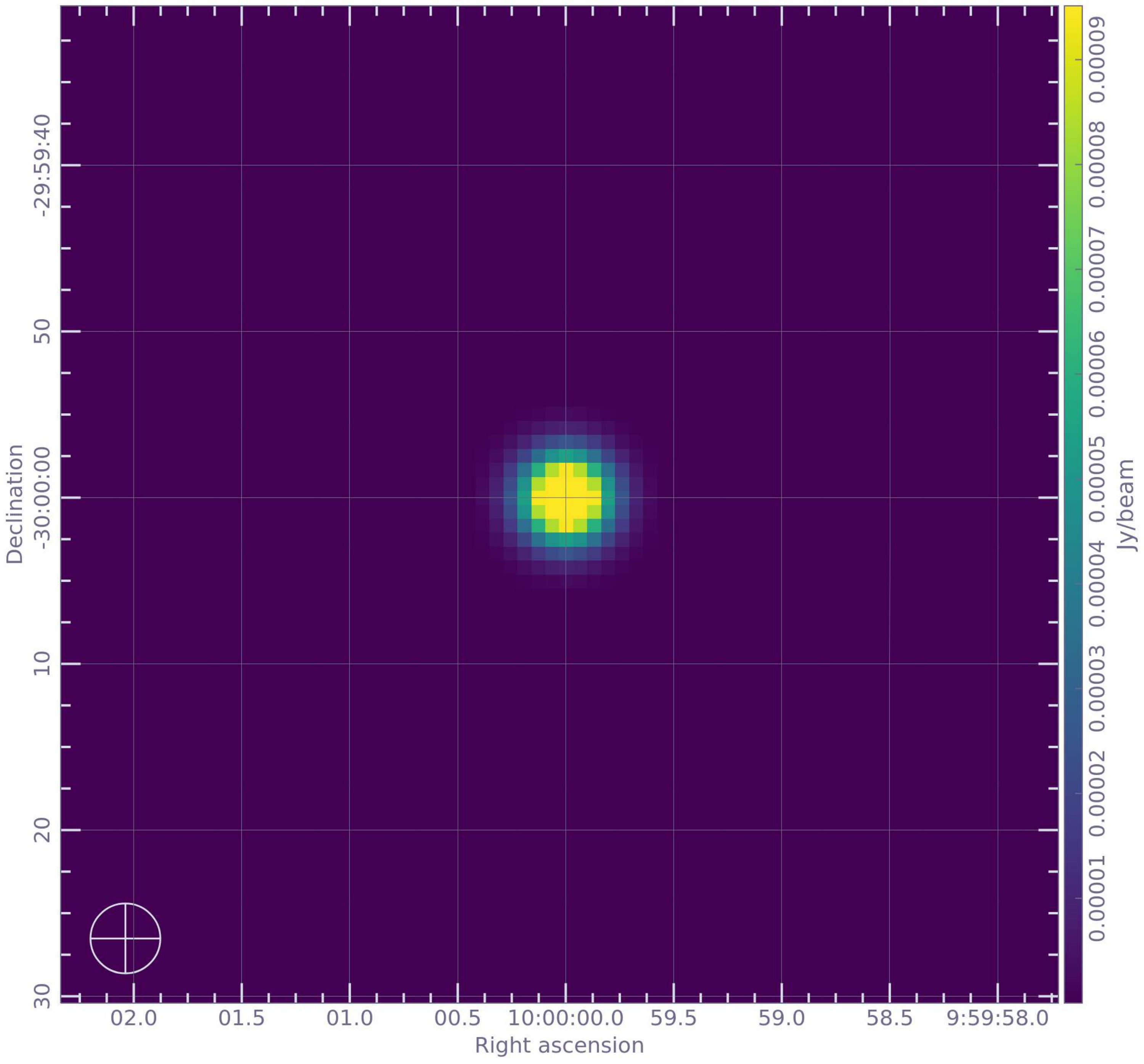}
    \includegraphics[trim={0.2cm 1.25cm 0.5cm 0cm},clip,width=0.24\linewidth]{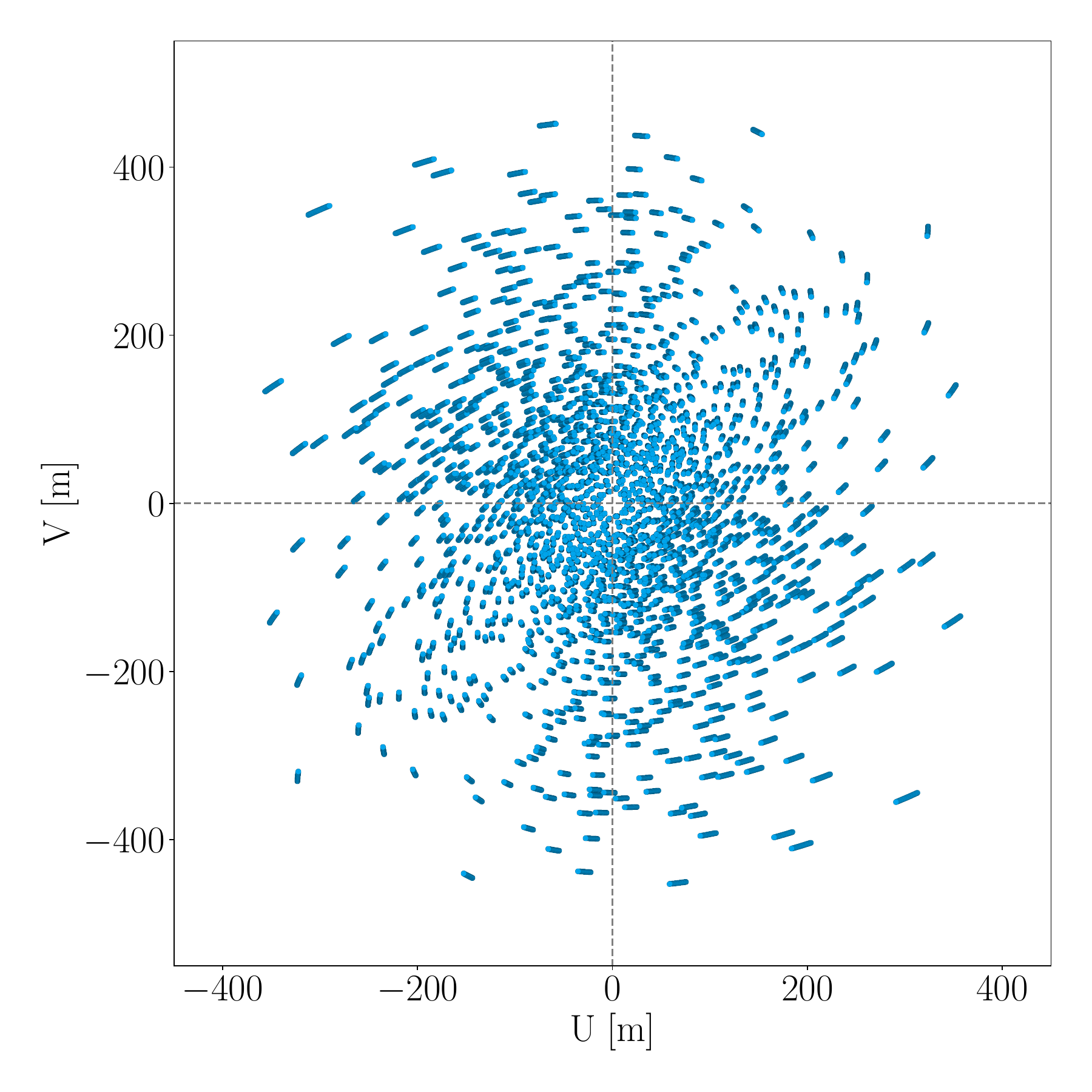}
        \includegraphics[trim={0.2cm 0cm 0.5cm 0.5cm},clip,width=0.24\linewidth]{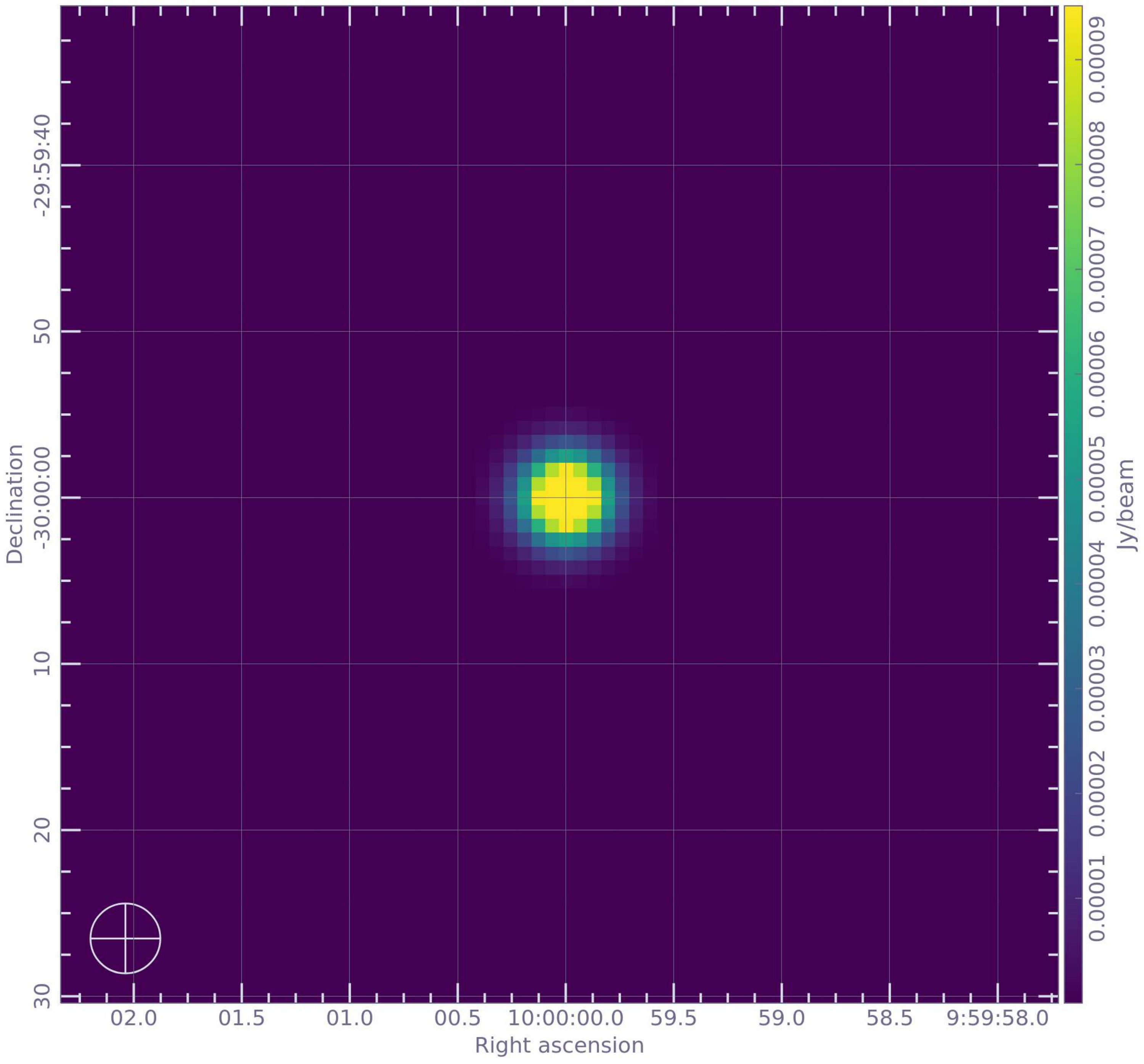}
    \includegraphics[trim={0.2cm 1.25cm 0.5cm 0cm},clip,width=0.24\linewidth]{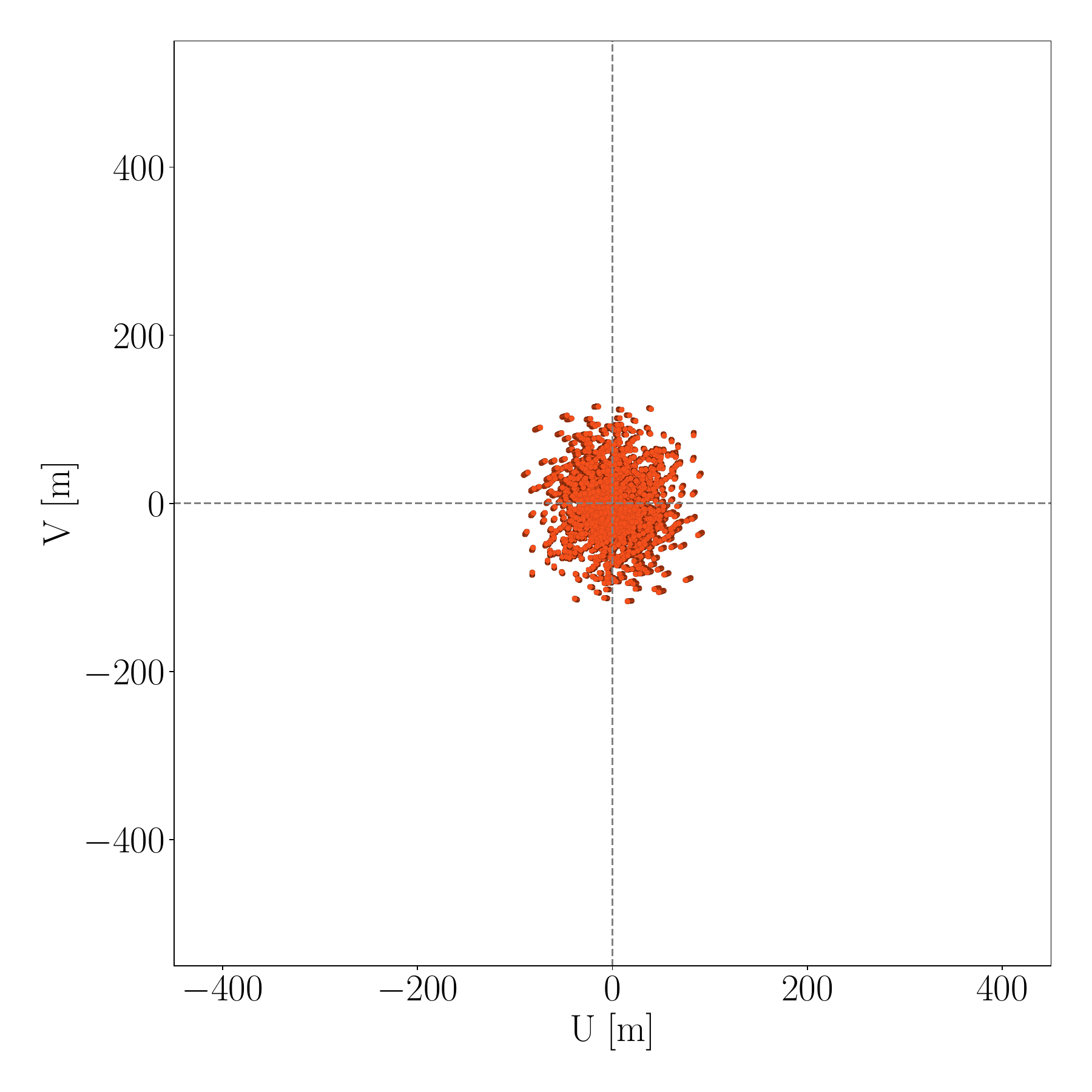}
    \caption{Example of the image and \emph{uv}-plane coverage in units of meters for one of the uncorrupted simulated sources at redshift 2.9, observed with ALMA configuration 4. The left panels (1 and 2) shows the image and \emph{uv}-plane coverage in the observed frame, while the right panels (3 and 4) shows the resulting image after rest-framing.}
    \label{fig:mocksource}
\end{figure*}

\begin{figure*}[!htpb]
    \centering
    \includegraphics[trim={0cm 0cm 0cm 0cm},clip,width=0.45\linewidth]{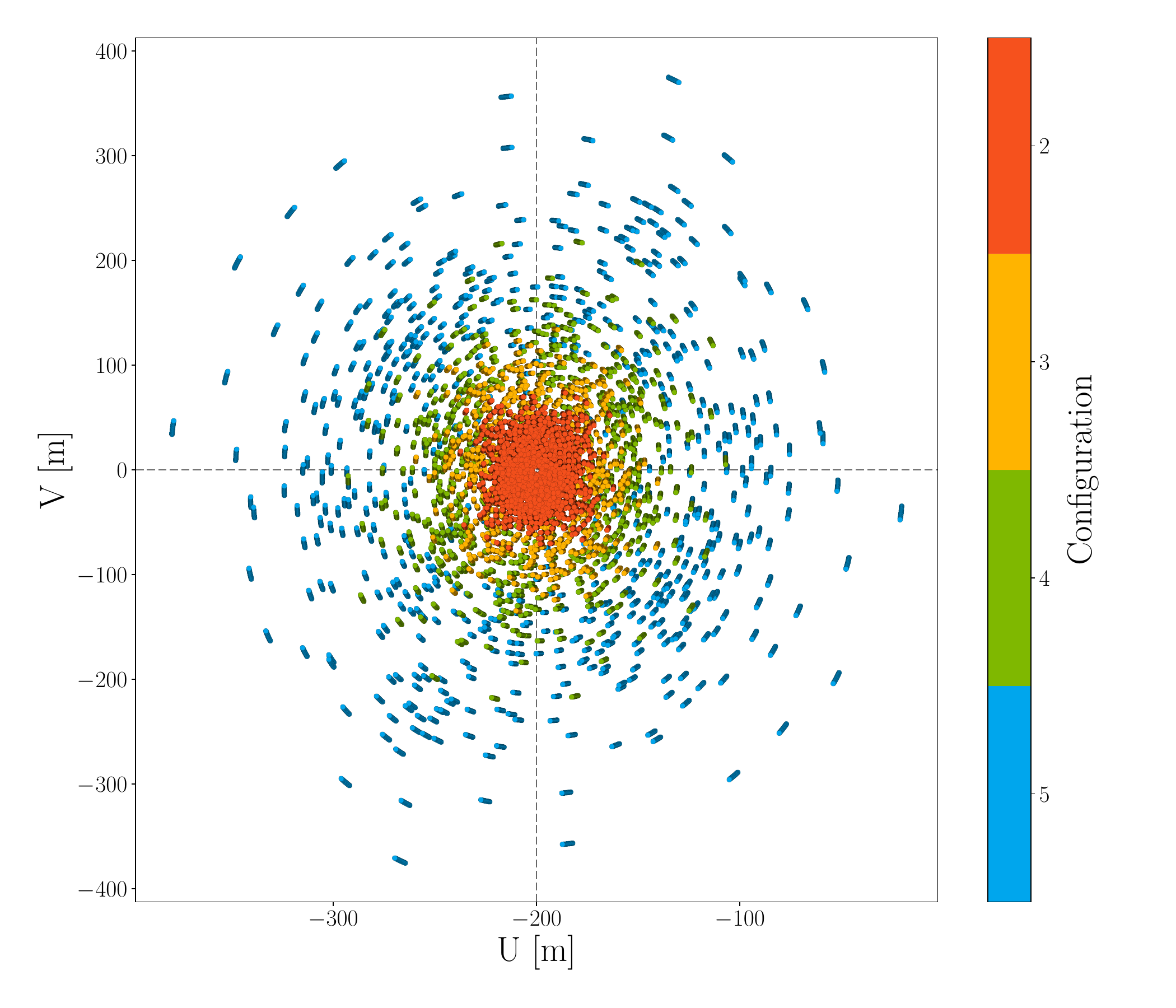}
    \includegraphics[trim={0cm 0cm 0cm 0cm},clip,width=0.45\linewidth]{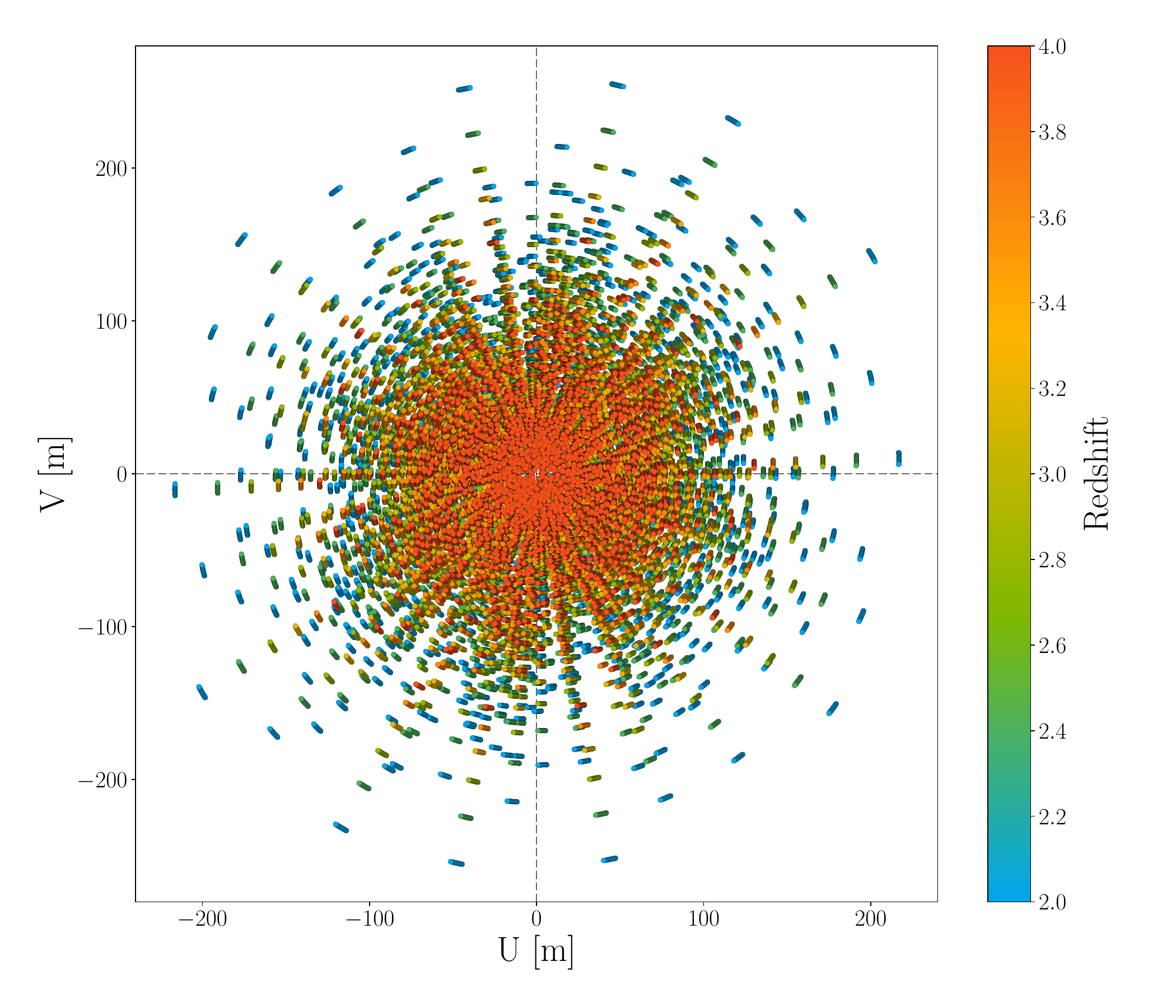}
    \caption{\emph{uv}-plane coverage of the stacked simulated source. Visibilities from different array configurations have been centered on the phase center and shifted to the rest frame using the \texttt{ViSta} method applied to the ALMA data. The left panel shows the rest-framed visibilities position in the \emph{uv}-plane for all the 50 mock observation, color-coded according to the observing configuration. The right panel shows the difference in \emph{uv}-plane coverage for the same configuration (5 sources at configuration 4, with max baseline at 780m ) observed at different redshifts.}
    \label{fig:combined_uv}
\end{figure*}

We generate simulations for two types of sources: a point-like source and an extended Gaussian source. For each case, we produce 50 synthetic observations of sources with redshifts randomly distributed between 2 and 4, using different antenna configurations from CASA’s predefined \texttt{alma.cycle.11.n.cfg} files, with $n$ ranging from 2 (corresponding to configurations with baselines up to 314 m) to 5 (with baselines up to 1400 m)\footnote{\url{https://almascience.eso.org/observing/observing-configuration-schedule}}.

The input sky models consist of spectral cubes with 100 channels, each 1 MHz wide. The reference frequency, corresponding to the central channel of all cubes, is set to 345.7598 GHz. This value is chosen to provide consistency with the real-case application of the method to the stacking of the CO(3–2) emission line, discussed later in the paper. However, it does not represent the line itself, and no specific molecular transition is modeled at this stage. All cubes are constructed such that their central channel corresponds to the observed frequency at which the redshifted line falls, ensuring that the same rest-frame frequency is aligned across sources, even though the observed frequencies differ due to their individual redshifts.
Each channel is represented by a 512$\times$512 pixel image, with a pixel scale of 0.05  $0.05^{\prime\prime}$, ensuring oversampling relative to the synthesized beam of all configurations. The sources are not placed exactly at the center of the image but are instead offset by a fixed amount from the central pixel, while still being fully contained within the sky model field of view.

In the point-source case, the cube includes a spatially and spectrally centered point-like source with both continuum and line emission. The continuum is modeled as a flat component with an intensity of $0.6$ mJy per channel across the full bandwidth. The line emission is added on top of the continuum and modeled as a Gaussian profile centered on the source position, confined to the 20 central spectral channels, with a peak flux of $2$ mJy. This observed frequency is chosen so that, at the considered redshifts, the emission line falls within ALMA Band 3. 

In the extended-source case, we model a spatially resolved Gaussian source with a FWHM of $5^{\prime\prime}$, also centered in space and frequency. This source includes the same flat continuum level ( $\sim 0.6$ mJy per channel) and the Gaussian profile centered on the middle channel and limited to the 20 central spectral channels, with a peak flux of $\sim 2$ mJy.

In both cases, we then corrupt the mock sources by adding the realistic thermal noise described above, following the functional form given by Equation (\ref{eq:noise}); for the redshift range considered in this study, the observed emission lines fall within ALMA Band 3, where the typical thermal noise level at a channel width of 1 MHz is about $\sim 1$ mJy \citep{nrao_casa_guides}.

These conditions ensure that the sources (both continuum and line emission) are not detectable in individual observations, enabling an exploration of detection and reconstruction strategies under low-SNR regimes.

For all sources, we avoided post-processing steps such as continuum subtraction to prevent bias in the characterization of the emission. Additionally, weighting was not applied since we considered only the frequency range common to all sources in the final analysis. We did not perform any correction for flux scaling due to redshift differences, as this is a simulated case where all fluxes were set equal from the beginning without any redshift-dependent adjustment.

Figure~\ref{fig:mocksource} provides a direct view of the effect of the \texttt{ViSta} rest-framing step described in Section~\ref{sec:restframe} on the \emph{uv}-plane of a single source. The \emph{uv}-plane coverage appears different, with the \emph{uv}-distances scaled down by a factor of 
$(1+z)$. Nevertheless, when the \texttt{ViSta} procedure is correctly applied and the MSs are properly adjusted, the resulting image is effectively identical to the original, differing only by the shift to the rest-frame frequency.

In Figure~\ref{fig:combined_uv}, we instead illustrate the \emph{uv}-plane coverage of the stacked mock observations after applying the \texttt{ViSta} procedure. The left panel displays the different array configurations adopted in the observing setups, while the right panel highlights the visibilities color-coded by redshift. The combination of a wide range of redshifts and multiple array configurations contributes to a denser and more homogeneous \emph{uv}-plane coverage. 

In the next section, we will present the results separately for unresolved (point-like) and resolved Gaussian sources. We will apply the \texttt{ViSta} method to fit the spectral profile of the sources, performing fits both in the visibility domain, which we will refer to as \texttt{ViSta}-\emph{uv}Fit, and in the image domain after imaging the stacked data, referred to as \texttt{ViSta}-ImFit. These results will then be compared with those obtained from the classical image-plane stacking procedure, which we will call \texttt{ImSta}-ImFit.

\subsection{Results} 
\subsubsection{Unresolved Source}

\begin{figure}[!htpb]
    \centering
    \includegraphics[trim={0.2cm 0cm 0cm 0cm},clip,width=0.49\linewidth]{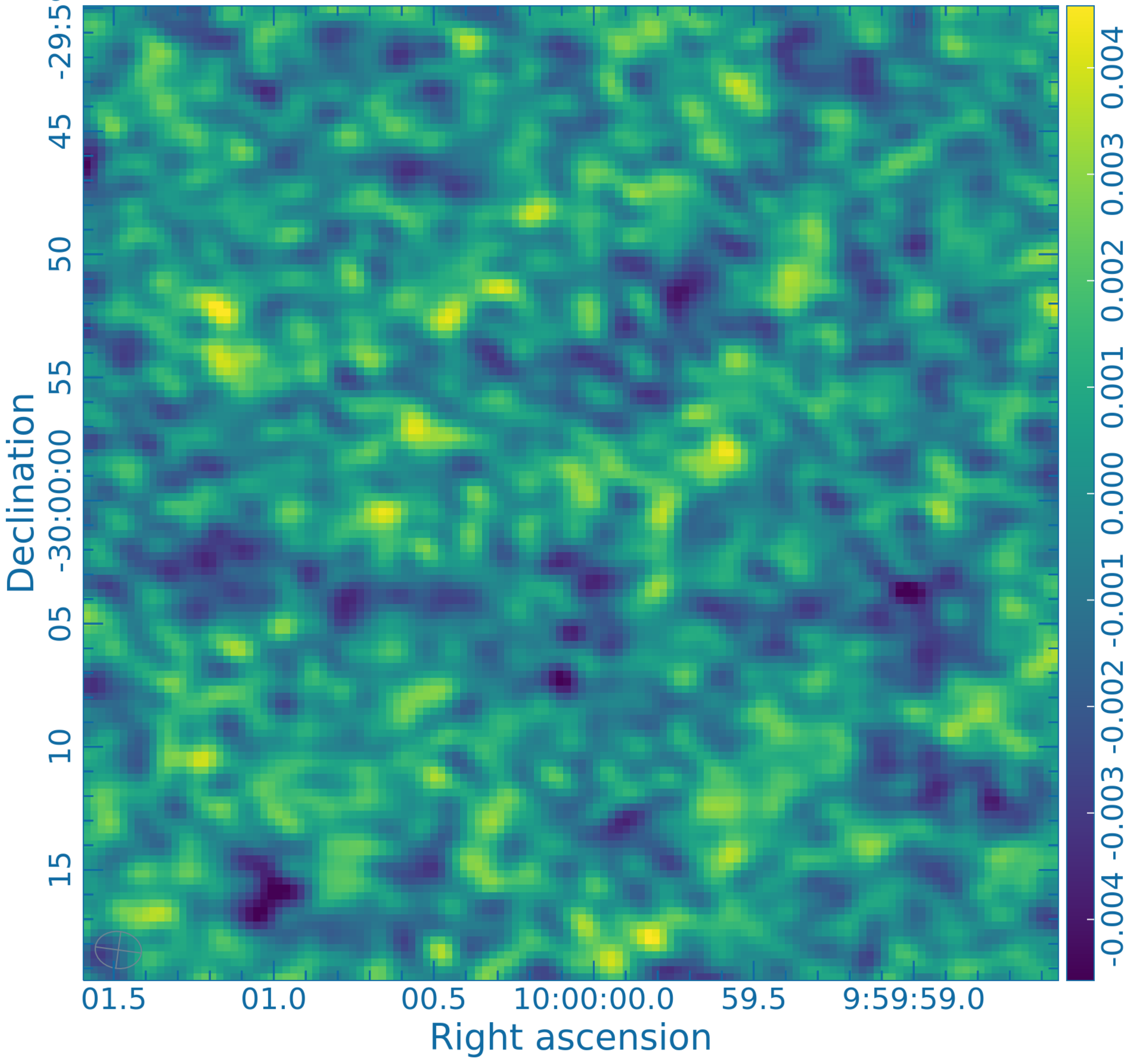}
    \includegraphics[trim={0.2cm 0cm 0cm 0cm},clip,width=0.49\linewidth]{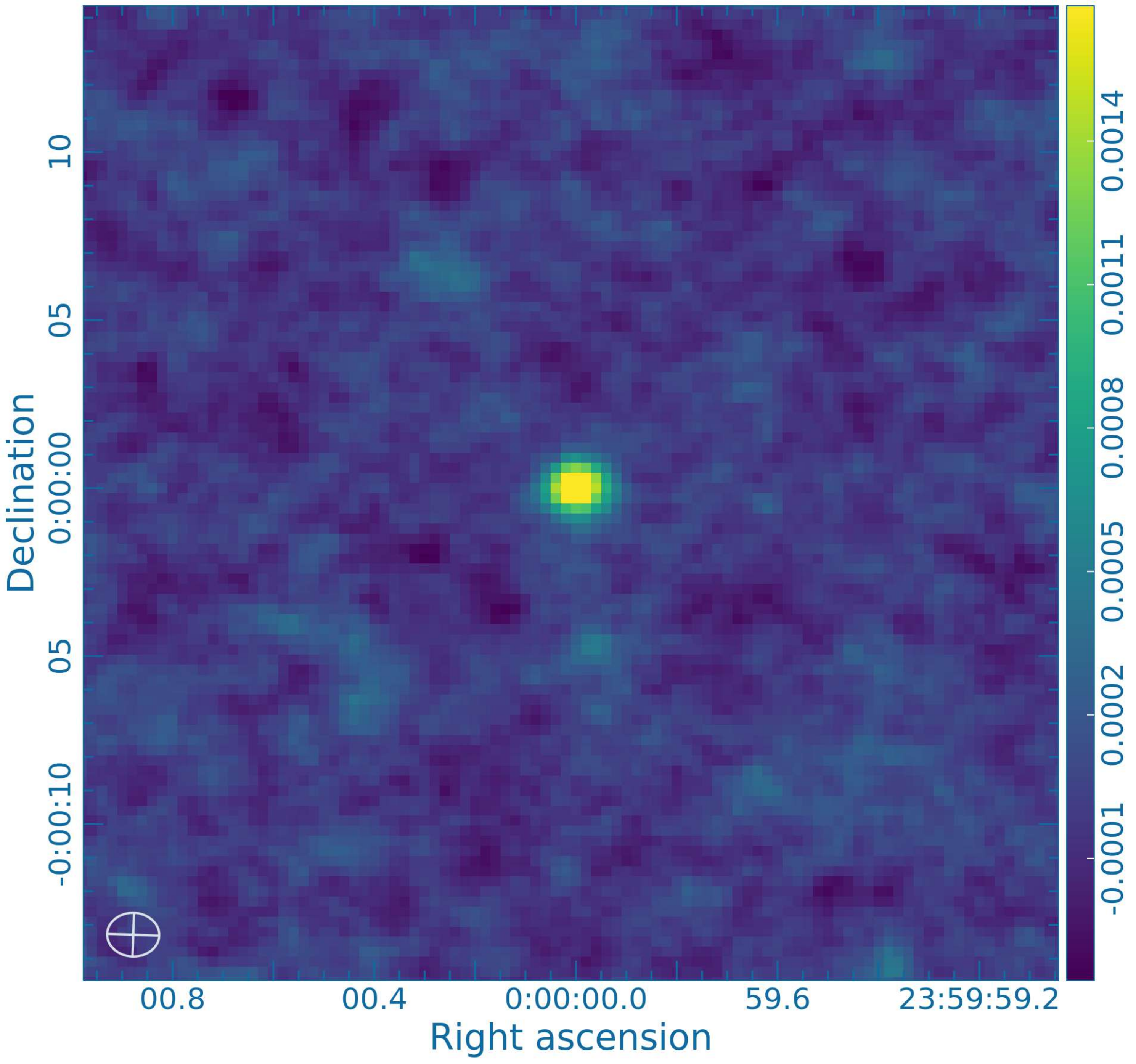}
    \caption{Comparison between the peak channel image of a single mock source before stacking, where the source signal is below the noise level (left panel), and the result after stacking 50 mock sources using the \texttt{ViSta} method (right panel).}
    \label{fig:single_vs_stacked}
\end{figure}

We analyse a sample of 50 simulated, perfectly point-like sources with the \texttt{ViSta} method. We begin with centering and the rest-framing of these sources (as discussed above) and then we proceed with the spectral rebinning and regridding onto a common frequency axis. The new channel width is determined by the source in the sample with the largest intrinsic frequency dispersion, which, since all mock sources initially share the same rest-frame properties, corresponds to the source at the highest redshift. We have performed a fit of the stacked data both in the \emph{uv}-plane and in the image plane. 

For the latter case, i.e. \texttt{ViSta}-Imfit, we follow the procedure described in \ref{sec:method}: we do not specify a restoring beam explicitly but we adopt the default one, i.e., a Gaussian fitted to the main lobe of the PSF resulting from the relatively uniform \emph{uv}-coverage of the combined dataset. Furthermore, we select natural weighting to enhance sensitivity at the expense of resolution; this choice is particularly suitable for sources that are essentially point-like at the observed resolution, similar to our case. Like we explained in \ref{sec:unresolved_sources}, for the case of point-like sources, we considered the peak value in the image as a measure of the flux density source, without integrating over an extended region.

\begin{figure}[!htpb]
    \centering
    \includegraphics[trim={1cm 0.5cm 1cm 1.2cm},clip,width=\linewidth]{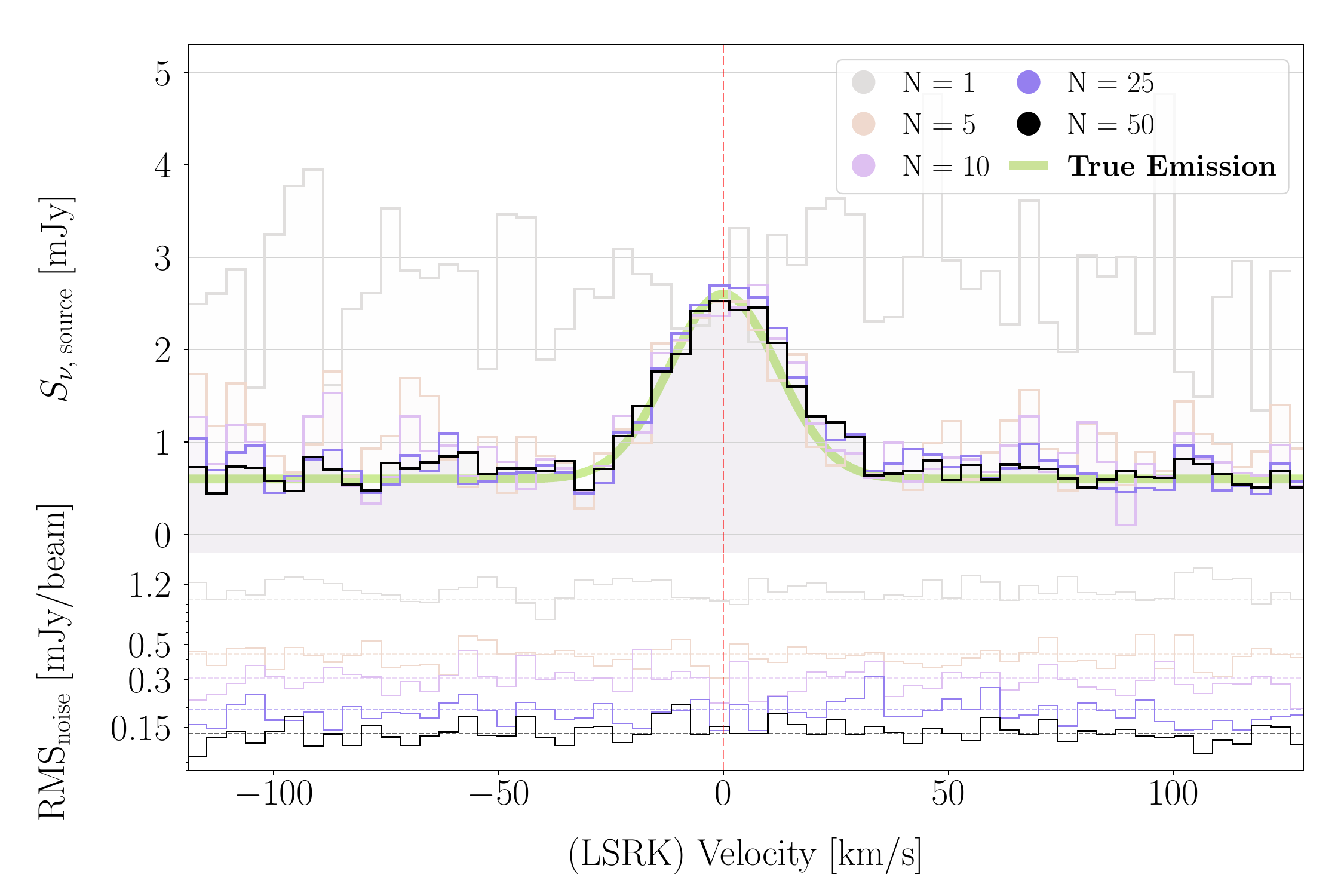}
    \caption{\textbf{Top panel:} Spectral line profiles recovered using the \texttt{ViSta}-\texttt{ImFit} method for different numbers of stacked point-like mock sources ($N = 1$, 5, 10, 25, 50), compared to the true emission profile (green solid line). \textbf{Bottom panel:} measured RMS noise of the corresponding stacked spectra as a function of $N$. The dashed line represents the expected $1/\sqrt{N}$ scaling, corresponding to the theoretical noise reduction with ideal stacking. The agreement between the measured and expected values confirms the statistical robustness of the method.}

    \label{fig:result_mock_sourcenumber}
\end{figure}

In Figure \ref{fig:single_vs_stacked}, we show the comparison between the image of a single mock source, where the signal is below the noise level, and the result obtained after stacking all 50 mock sources using the \texttt{ViSta} method. As can be seen, while the individual source is not detectable, the stacking procedure clearly amplifies the signal. Figure \ref{fig:result_mock_sourcenumber} further illustrates how the signal-to-noise ratio improves as a function of the number of stacked sources, and the corresponding quantitative measurements for the central channel are reported in Table \ref{tab:sim_imSta}.
The advantage of the \texttt{ViSta} method increases with the number of stacked sources. As expected, the \emph{rms} noise decreases approximately as $1/\sqrt{N}$ (the trend illustrated by the dashed line in Figure~\ref{fig:result_mock_sourcenumber}). In our case, stacking 50 datasets reduces the average channel \emph{rms} $\sigma_{\rm chan}^*$ (initially $\sim1.2$ mJy for the reference image) by a factor $\sim 8$ consistent with this scaling, giving a final peak SNR of approximately $17.3$ on stacking.

\begin{deluxetable}{cccc}[!htpb]
\tabletypesize{\scriptsize}
\tablewidth{0pt}
\tablecaption{Peak SNR Results for \texttt{ViSta}-ImFit}
\tablehead{
\colhead{$N_{\text{Source}}$} & 
\colhead{$S_{\text{peak}}$ (mJy/beam)} & 
\colhead{$\sigma_{\text{chan}}$ (mJy/beam)} & 
\colhead{Peak SNR}
}
\startdata
1  & 3.3125 & 1.1191 & 2.9600 \\
5  & 2.5168 & 0.4240 & 5.9358 \\
10 & 2.4579 & 0.3009 & 8.1695 \\
25 & 2.6655 & 0.1876 & 14.2058 \\
50 & 2.4283 & 0.1404 & 17.3007 \\
\enddata
\tablecomments{Signal-to-noise ratio (SNR) results for the \texttt{ViSta}-ImFit procedure for the central channel. $S_{\text{peak}}$ is the peak flux, $\sigma_{\text{chan}}$ is the average channel \emph{rms}.}
\label{tab:sim_imSta}
\end{deluxetable}

\begin{figure}[!htpb]
    \centering
    \includegraphics[trim={1.3cm 0.5cm 1cm 0.8cm},clip,width=\linewidth]{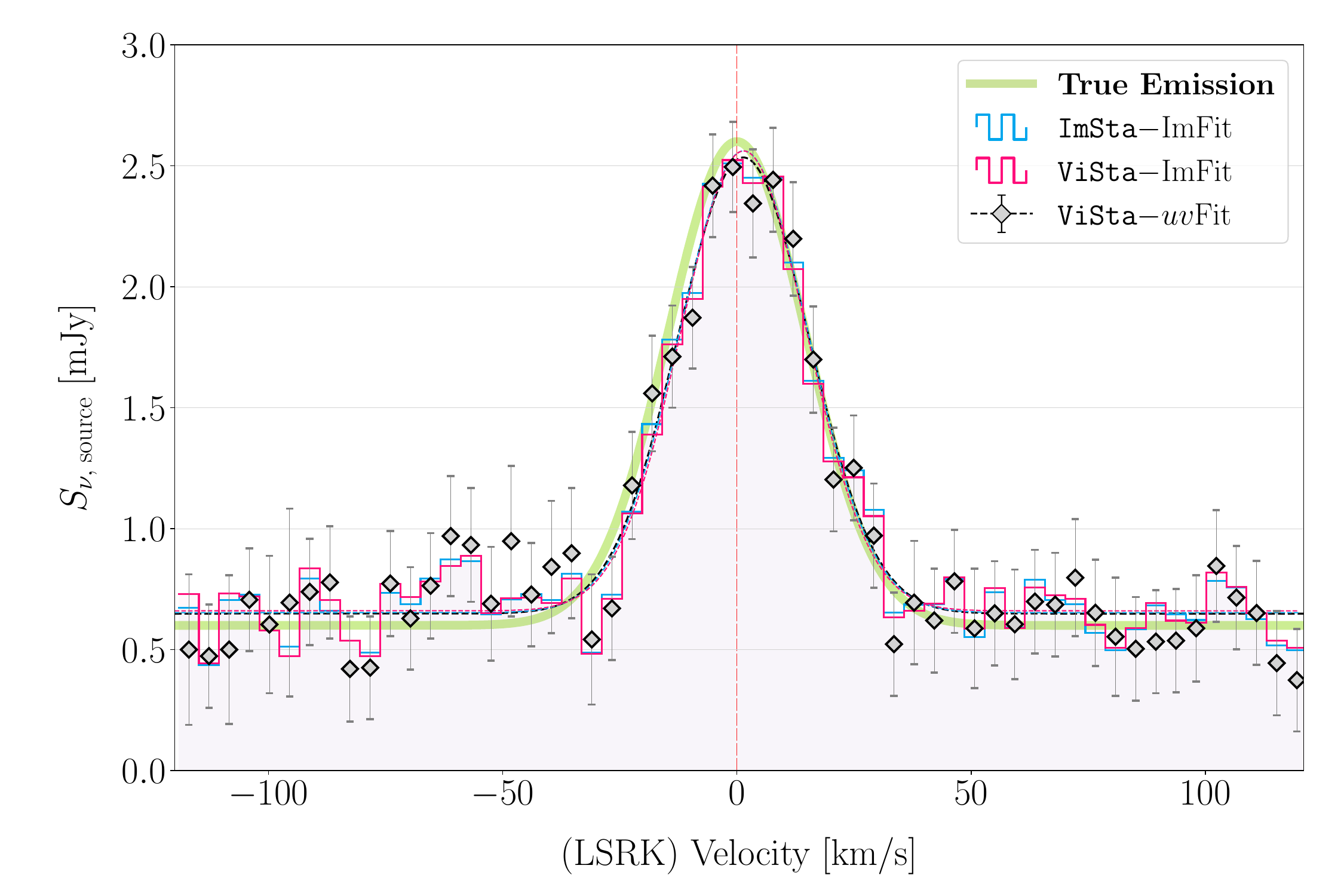 }
    \caption{Comparison of the recovered spectral line profiles obtained using different stacking and fitting methods on simulated data for unresolved sources. The green solid line represents the true emission model used in the simulation. The blue and the magenta lines show respectively the results from the \texttt{ImSta}-ImFit and  \texttt{ViSta}-\emph{uv}Fit procedures, and the dashed lines represent the Gaussian fits to the data points. The diamonds represent the data points fitted in the \texttt{uvFit} method with 1$\sigma$ uncertainties; for the  \texttt{ViSta}-\emph{uv}Fit case, where error bars were available, the Gaussian fit was performed via Monte Carlo sampling: synthetic realizations of the data were generated within the uncertainties, and the final fit (black dashed line) represents the average result. The plot shows that the three methods yield consistent results and recover the signal reasonably well.}

    \label{fig:result_mock_images}
\end{figure}

We then performed a spectral fitting directly in the visibility domain using \texttt{uvm.uvmultifit}, adopting a Gaussian model to provide greater flexibility in the source representation. To assess the consistency of this method, we also compared it with the standard image-stacking approach \texttt{ImSta}-ImFit. For each target in the sample, we reconstructed the image using its original observational parameters, i.e. synthesized beam, sensitivity and field of view. Following a procedure analogous to the \texttt{ViSta}-ImFit, we extracted the flux density from the central pixel in each spectral channel. The final stacked spectrum was then obtained by averaging the flux values across all sources for each channel.

Figure \ref{fig:result_mock_images} presents the stacked spectra obtained using the various methods described above. The green line represents the expected average emission, derived from averaging all uncorrupted observations. The black diamonds indicate the best-fit flux densities from the \texttt{ViSta}-\emph{uv}Fit procedure, with corresponding uncertainties, while the black dashed line represents the Gaussian model fitted to these points. The fit was obtained through a Monte Carlo approach, by perturbing the data according to their uncertainties and averaging the resulting fits. The magenta and blue lines correspond to the spectral fits obtained using the \texttt{ViSta}-ImFit and \texttt{ImSta}-ImFit methods, respectively, with the dashed lines in matching colors indicating the corresponding Gaussian fits.
All three methods successfully reproduce the Gaussian line profile, recovering the true emission with about 90\% accuracy and showing only minor differences among them. However, they appear to slightly overestimate the continuum level. This discrepancy may arise because the continuum is significantly weaker and comparable to the average noise level, making it more difficult to constrain reliably. These results indicate that all three approaches are robust and fully reliable for the analysis of unresolved sources.

\subsubsection{Resolved Source}\label{sec:nativesize}
\begin{figure}[!htpb]
    \centering
    \includegraphics[trim={1cm 0.5cm 1cm 1.2cm},clip,width=\linewidth]{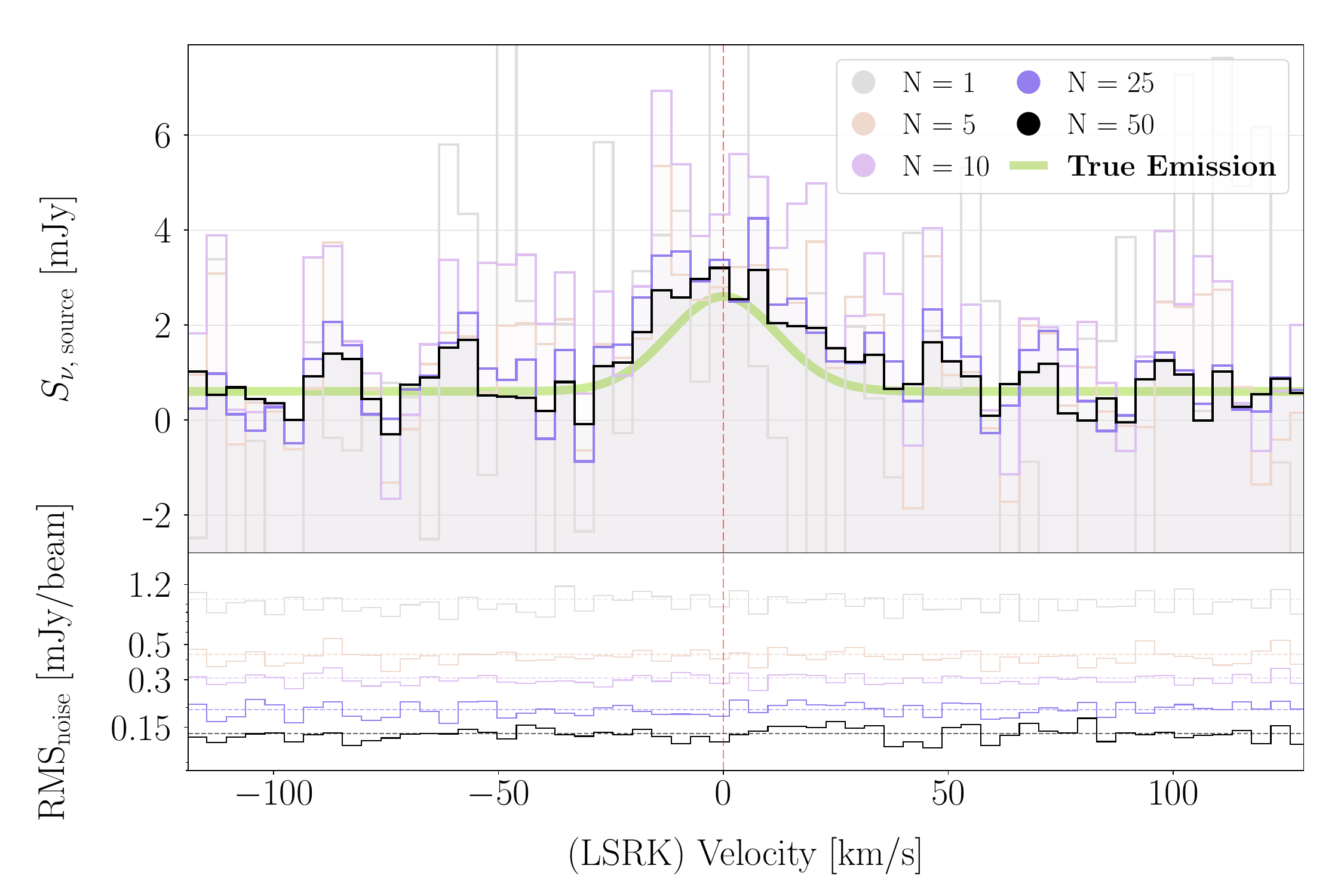}
    \caption{\textbf{Top panel:} Spectral line profiles recovered using the \texttt{ViSta}-\texttt{ImFit} method for different numbers of stacked mock extended sources ($N = 1$, 5, 10, 25, 50), compared to the true emission profile (green solid line). \textbf{Bottom panel:} measured RMS noise of the corresponding stacked spectra as a function of $N$. The dashed line represents the expected $1/\sqrt{N}$ scaling, corresponding to the theoretical noise reduction with ideal stacking. The agreement between the measured and expected values confirms the statistical robustness of the method.}
    \label{fig:scaling_gauss}
\end{figure}

For resolved or partially resolved sources, the stacking procedure remains conceptually identical to that adopted for point-like sources; the main difference lies in how the flux density is measured. In this regime, the peak pixel value no longer provides a reliable estimate of the total emission, and the flux must instead be integrated over a spatially extended region.
To explore this, we present an analysis based on three different scenarios, always comparing our method to the standard \texttt{Vista}-Imfit stacking approach. In the first case, we assume that the intrinsic size of the sources is known, allowing us to measure the flux density directly within the corresponding area. In the second case, we assume no prior knowledge of the source size, and instead fit each detection with a 2D Gaussian model using \texttt{castasks.imfit()}, from which we derive both the morphology and the integrated flux. In the third case, where the individual sources are not suitable for reliable fitting, we convolve all images to a common (i.e., lowest) spatial resolution to ensure consistency. The stacking is then performed on the convolved images, followed by a fit on the stacked result.

\paragraph{Native Source Size at Native Resolution}

\begin{figure}[!htpb]
    \centering
    \includegraphics[trim={2cm 0.5cm 1cm 1cm},clip,width=\linewidth]{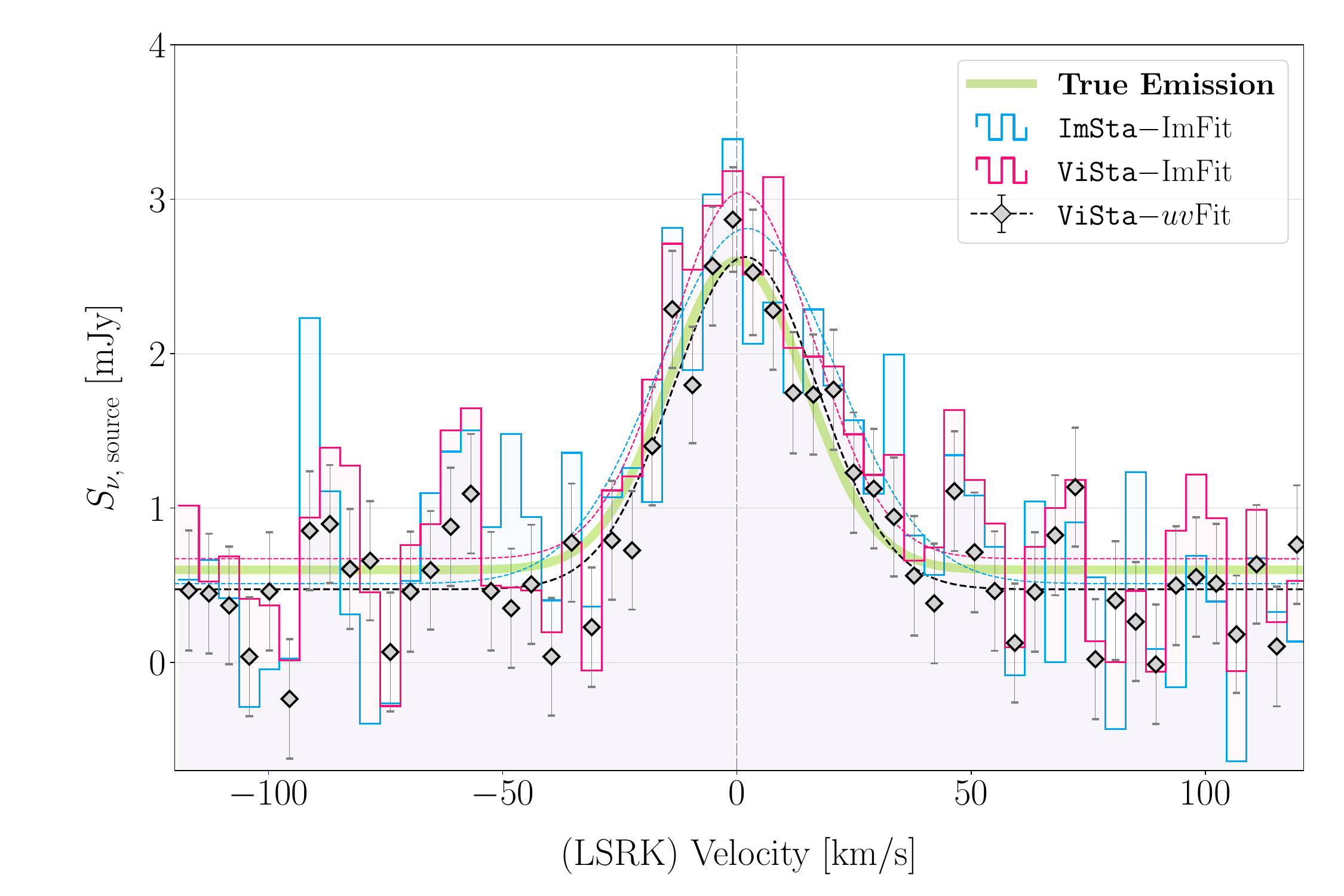}
    \caption{Comparison of the recovered spectral line profiles obtained using different stacking and fitting methods on simulated data for a resolved source, for known source sizes. The green solid line represents the true emission model used in the simulation. The blue and  magenta lines show respectively the results from the \texttt{ImSta}-ImFit and \texttt{ViSta}-ImFit procedures, and the dashed lines represent the Gaussian fits to the data points. The diamonds represent the data points fitted in the \texttt{ViSta}-\emph{uv}Fit method with 1$\sigma$ uncertainties and the dashed lines is the Gaussian fit performed via Monte Carlo sampling.}
    \label{fig:gauss_orig_size}
\end{figure}

The first test focuses on the ideal case in which the intrinsic size of the sources is well known. This option suits our simulated case well, where all sources share the same spatial extent, characterized by a FWHM of $5^{\prime\prime}$. This allows us to measure the flux density directly within the appropriate region, using a circular aperture with a radius of 3$\sigma$ (where $\sigma$ is the standard deviation of the Gaussian profile) to encompass the majority of the source emission while minimizing contamination from the background noise. For the \texttt{Vista}-ImFit, after the classic pre-processing and stacking procedure, we performed the imaging step as per standard procedure and we measured the flux density for every channel of the stacked cube within the known source extension. For the \texttt{Vista}-\emph{uv}Fit, we use the Gaussian model within the fixed region, letting only the embedded flux change. Finally, for the image stacking  \texttt{Imsta}-ImFit, we generate every image for every mock observation like for the point-like scenario, we extract for each source the flux density embedded in the same region and we then average the obtained values along the spectral axis.

In this context, we also investigated how the performance of the \texttt{ViSta}-ImFit method evolves with the number of individually extracted sources included in the stack and we present the results in Figure \ref{fig:scaling_gauss}. Similar to the point-source case, the noise decreases as expected, scaling approximately with the square root of the number of sources. However, we find that recovering a well-defined spectral line profile requires a significantly larger sample. This is because, when integrating over an extended area, a greater amount of noise is also accumulated, reducing the efficiency with which the spectral signal emerges. As a result, although the method remains effective, the convergence toward the intrinsic line profile is slower than in the unresolved case, and a larger number of sources is needed to achieve comparable spectral quality.

The stacked spectra resulting from the different methods are shown in Figure \ref{fig:gauss_orig_size}, using the same color map and linestyles used for the point-source case. All three approaches recover the line emission with good fidelity. The direct visibility‑plane extraction via \texttt{Vista}-\emph{uv}Fit provides the closest match to the expected profile, whereas  \texttt{Vista}-ImFit tends to slightly overestimate the Gaussian peak intensity. This problem can be attributed to the heterogeneous beam configurations of the individual datasets. When these observations are concatenated, the resulting effective beam differs subtly from each input synthesized beam, which can concentrate flux into a narrower core and artificially boost the measured peak. Consequently, slight variations in beam size and shape across the sample propagate into the stacked spectrum, leading to the observed bias of the stacked source size.
Finally, the spectra derived from \texttt{Imsta}-ImFit exhibit larger point‑to‑point scatter, likely a consequence of including more noise in the integration region.

\paragraph{Estimated Source Size at Native Resolution}\label{sec:estimate_nativeres}

\begin{figure}[!htpb]
    \centering
    \includegraphics[trim={2cm 0.5cm 1.05cm 1cm},clip,width=\linewidth]{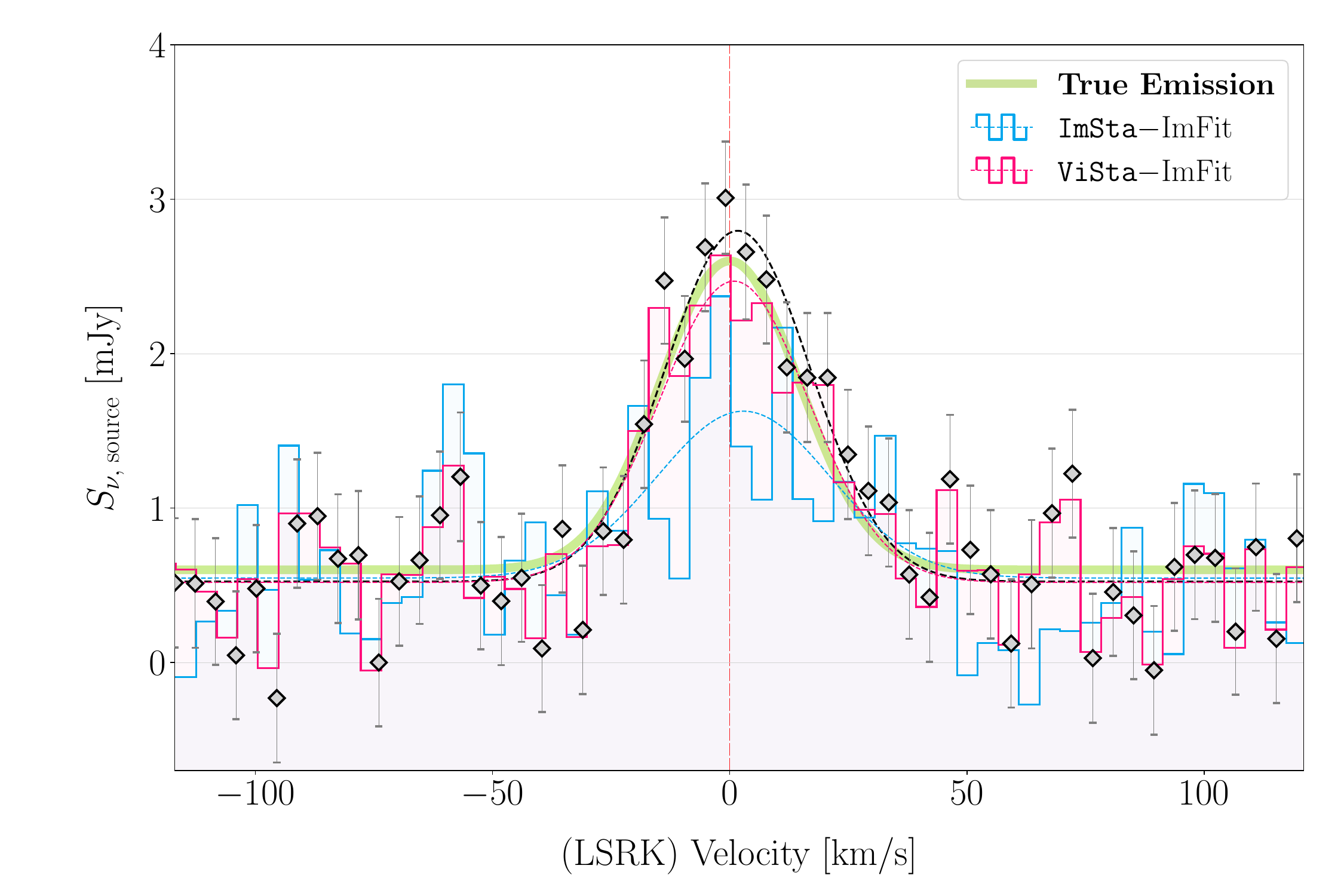}
    \caption{Comparison of the recovered spectral line profiles obtained using different stacking and fitting methods on simulated data for a resolved source, derived after fitting the size of the source. The green solid line represents the true emission model used in the simulation. The blue and the magenta lines show respectively the results from the \texttt{ImSta}-ImFit and \texttt{ViSta}-ImFit procedures, and the dashed lines represent the Gaussian fits to the data points. The diamonds represent the data points fitted in the \texttt{ViSta}-\emph{uv}Fit method with 1$\sigma$ uncertainties and the dashed lines is the Gaussian fit performed via Monte Carlo sampling.}
    \label{fig:gauss_imfit}
\end{figure}

We repeat the analysis presented above without assuming prior knowledge of the source size. In the \texttt{ViSTa}–\emph{uv}Fit approach, to obtain a more accurate estimate of the source geometry, we first performed a fit on the data averaged over all channels. The best-fit size parameters (major, minor axes and orientation of the spatial Gaussian) were then kept fixed and used to estimate the flux density independently in each individual channel. While fitting in the image plane, on the other hand, we use CASA’s \texttt{imfit} task: it performs a two‑dimensional Gaussian fit to a user‑defined region of emission, deconvolving the trial model from the synthesized beam to recover the intrinsic source morphology. \texttt{imfit} returns several best‐fit parameters : integrated flux density, centroid coordinates, deconvolved major and minor axes, and position angle. 
For the \texttt{ViSTa}–ImFit workflow, we first perform the uv‑stacking and then compute the zero‑moment map of the stacked visibilities. We then fitted the 2D Gaussian model to the moment‑0 image, yielding the deconvolved source size. In contrast, in the (\texttt{ImSta}–ImFit) approach, we begin by generating a moment‑0 map for each individual cube. We then fit each map with a 2D Gaussian, extract the integrated flux density within the fitted aperture for every source, and finally average these per‑source fluxes across the entire sample on a channel‑by‑channel basis to assemble the stacked spectrum. For both methods, the flux density was extracted within a radius of \(3\sigma\) from the fitted Gaussian. 
The results are presented in Figure \ref{fig:gauss_imfit}. They highlight the robustness of the \texttt{ViSTa} method in recovering both continuum and line emission, whether the fitting is performed in the \emph{uv} plane or in the image plane. This is largely due to its ability to accurately resolve the source size, even when the emission is faint and spatially extended, allowing for a clean flux measurement that is minimally affected by noise. On the contrary, the \texttt{ImSta}–ImFit approach struggles to precisely determine the intrinsic source size of single sources, even in the continuum map. This leads to inaccurate flux estimates for the line emission and results in a stacked spectrum that shows a lower peak emission and significantly larger fluctuations.

\paragraph{Estimated Source Size at Lowest Resolution}\label{sec:lowest_res}

\begin{figure}[!htpb]
    \centering
    \includegraphics[trim={1.3cm 0.5cm 1.05cm 1cm},clip,width=0.96\linewidth]{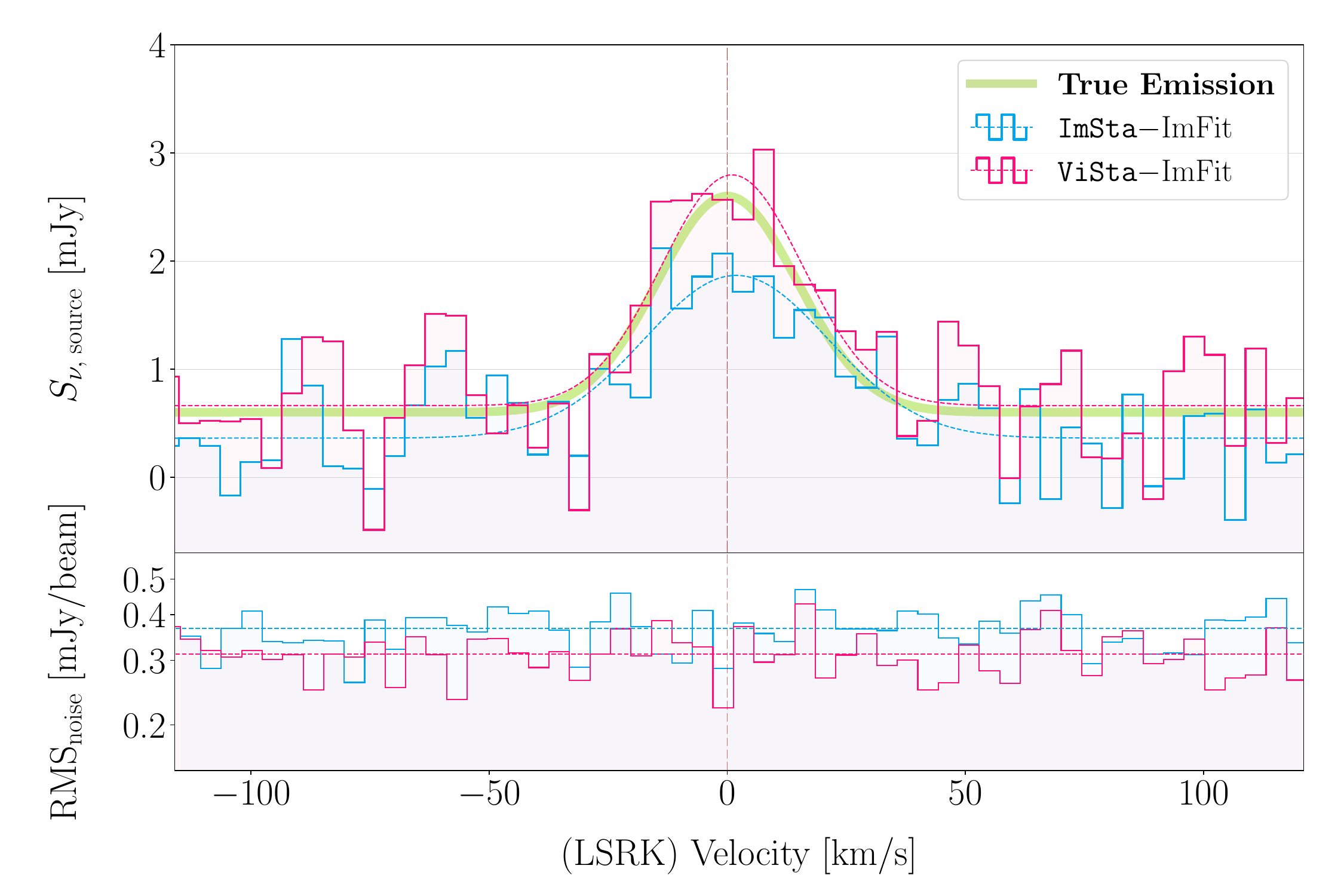}
    \caption{Comparison of the recovered spectral line profiles obtained using different stacking and fitting methods on simulated data for resolved source, after convolving all sample to the lowest resolution. The green solid line represents the true emission model used in the simulation. The blue and the magenta lines show respectively the results from the \texttt{ImSta}-ImFit and \texttt{ViSta}-ImFit procedures, and the dashed lines represent the Gaussian fits to the data points.}
    \label{fig:lowest_res_results}
\end{figure}

\begin{figure}[!htpb]
    \centering
    \includegraphics[trim={0.4cm 0.4cm 0cm 0cm},clip,width=0.49\linewidth]{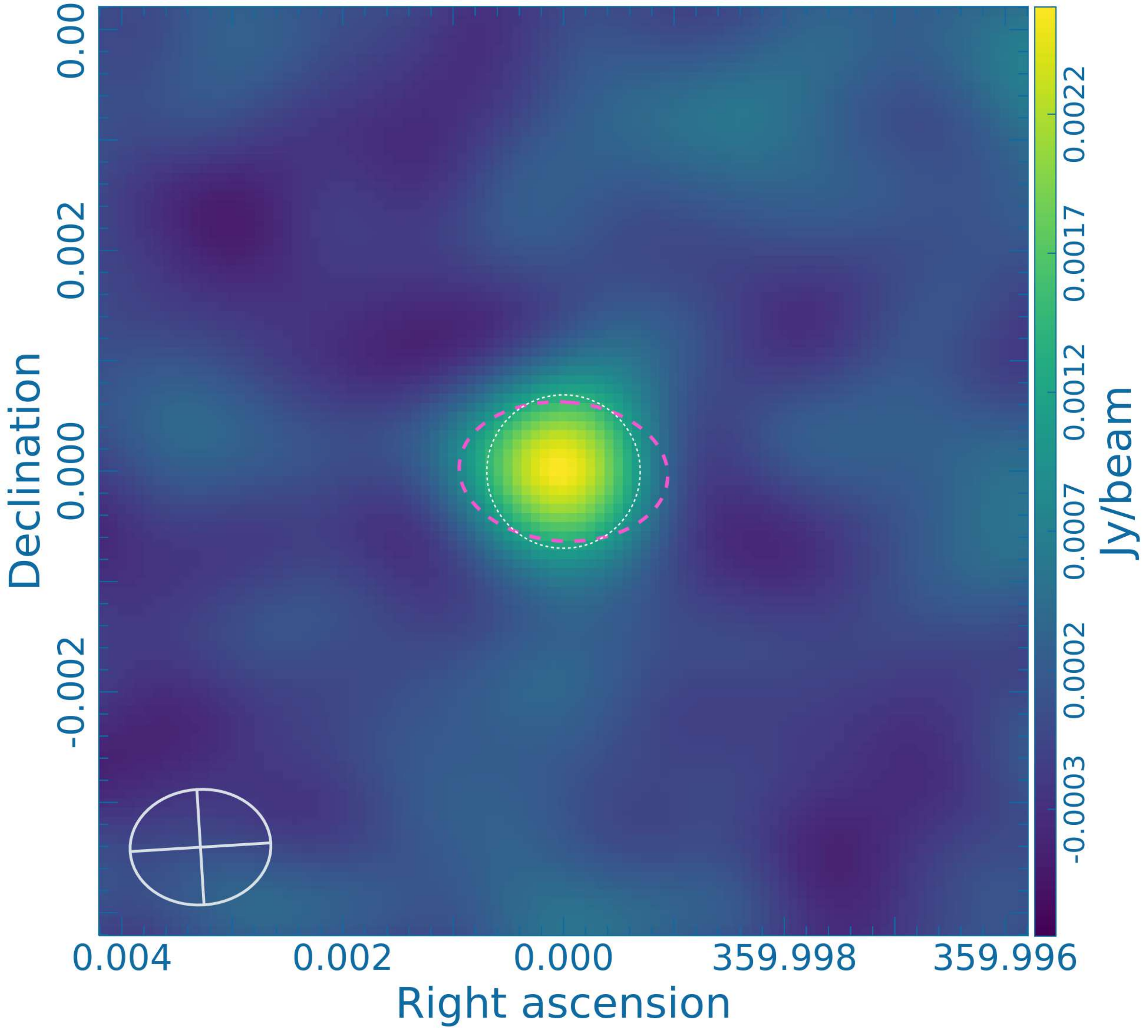}
    \includegraphics[trim={0.4cm 0.4cm 0cm 0cm},clip,width=0.49\linewidth]{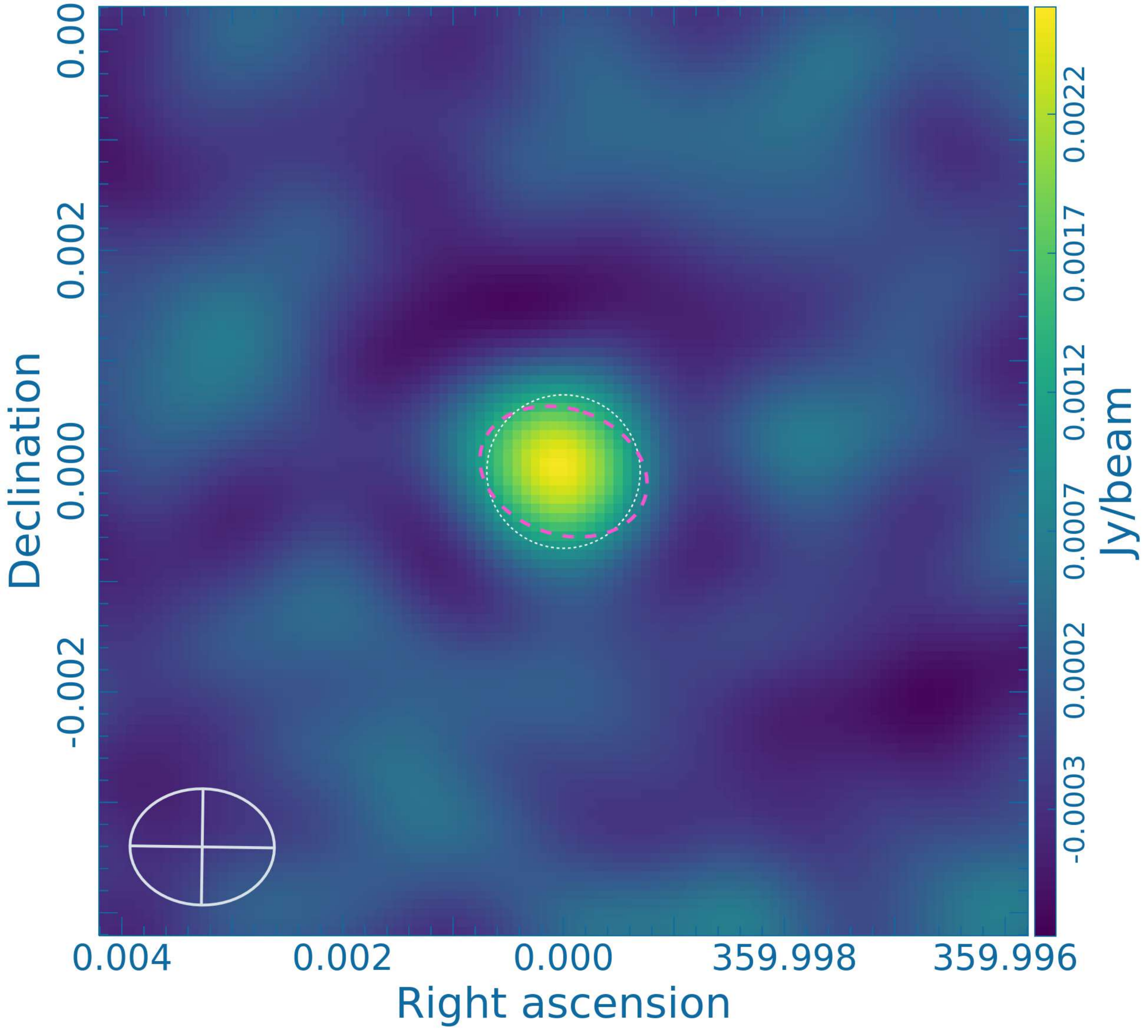}
    \caption{Stacked images of the central channel of the cube using two different convolution methods. \textbf{Left:} Image stacked in the image plane using the \texttt{ViSTa} method, then convolved with a largest beam of the sample. \textbf{Right:} Image stacked in the image plane after convolving all input images to the worst resolution. The dashed purple circle indicates the source size as fitted with \texttt{casatasks.imfit()}, while the white dotted circle shows the effective size of the stacked source. The intrinsic source size is reliably recovered by both methods, but visibility-plane stacking yields a smoother background.}
    \label{fig:fig2_conv}
\end{figure}

Finally, we repeated the analysis on the full sample after degrading all data to the lowest common angular resolution. In this configuration, both the integrated flux density and the noise level (now uniform across all images in the dataset) were estimated using the (\texttt{ViSTa}–ImFit) and (\texttt{ImSTa}–ImFit) approaches. Figure \ref{fig:lowest_res_results} shows the results of both image‑plane fitting methods applied to the two stacking strategies at the lowest common resolution. In this case, stacking directly in the \emph{uv} domain results in a more uniform filling of the Fourier plane and naturally down-weights the noisiest baselines, yielding a higher-fidelity reconstruction of the source core. In contrast, convolving each individual image to the lowest resolution, each with relatively low SNR, introduces additional noise into stacking, leading to a broader scatter in the measured spectrum, a reduced peak amplitude, and a slightly elevated noise. Thus, after resolution matching, when the SNR is low in the single observation of the sample, the \texttt{ViSTa} method consistently better captures the true source profile.
Figure \ref{fig:fig2_conv} shows the convolved stacked image with both methods for the central channel of the cube. In addition, it also demonstrates that the deconvolved source size recovered in both the \texttt{ViSTa} and \texttt{ImSTa} methods closely matches the true input value. However, the image-plane extraction exhibits a more pronounced noise pattern, with an average RMS for the central channel of approximately $\sim0.285$ mJy/beam, whereas the one stacked in the visibility plane yields a smoother background, with an RMS of $\sim0.22$ mJy/beam. Considering the flux densities obtained with these two methods, respectively $2.070$ and $2.566$ mJy, the \texttt{ViSTa} method yields a gain of $\sim54\%$ in the central channel compared to the image-plane stacking.

\paragraph{Continuum extraction at lowest resolution}

\begin{figure}[!htpb]
    \centering
    \includegraphics[trim={0.3cm 0.3cm 0.2cm 0cm},clip,width=0.49\linewidth]{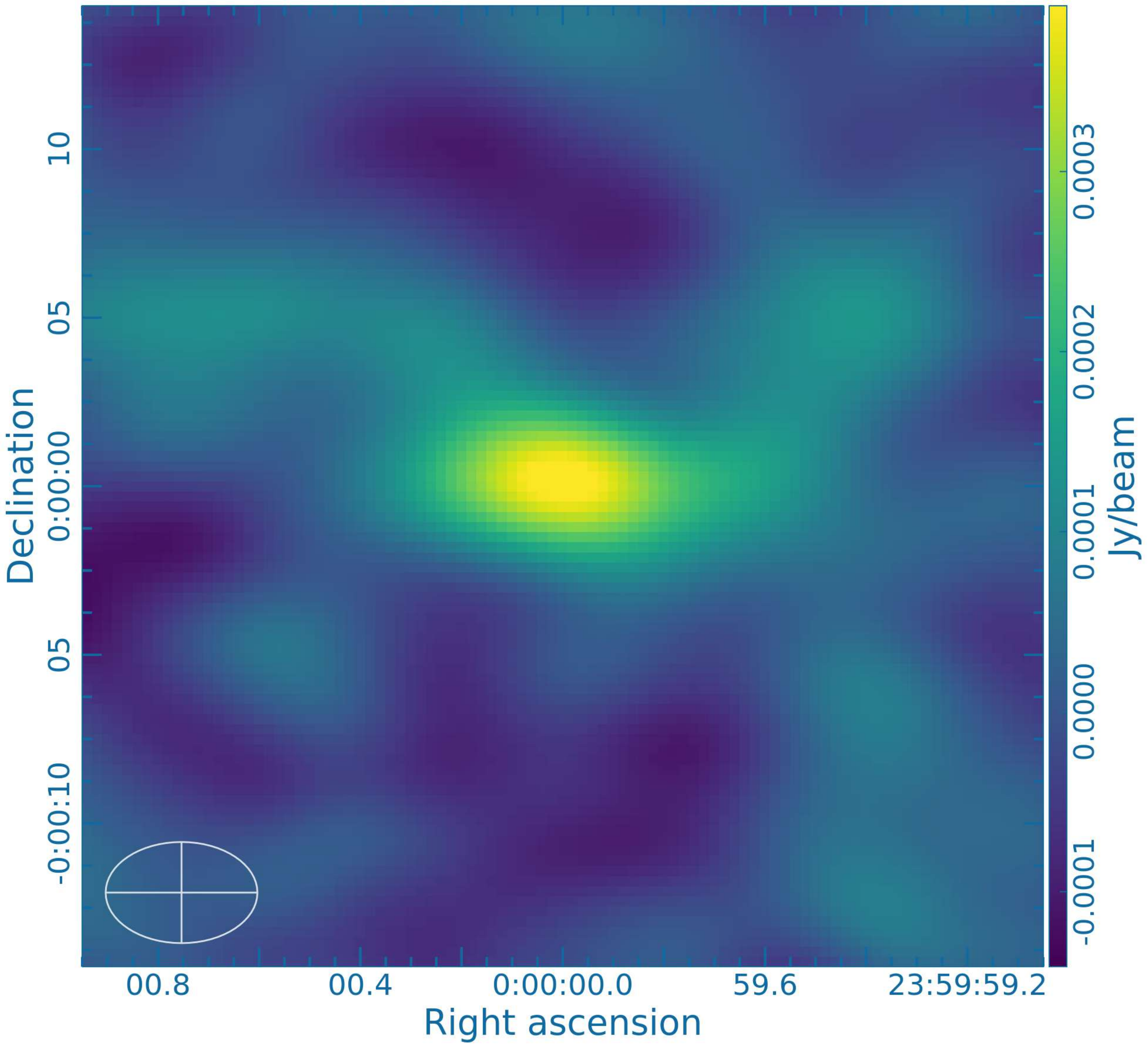}
    \includegraphics[trim={0.3cm 0.3cm 0.2cm 0cm},clip,width=0.49\linewidth]{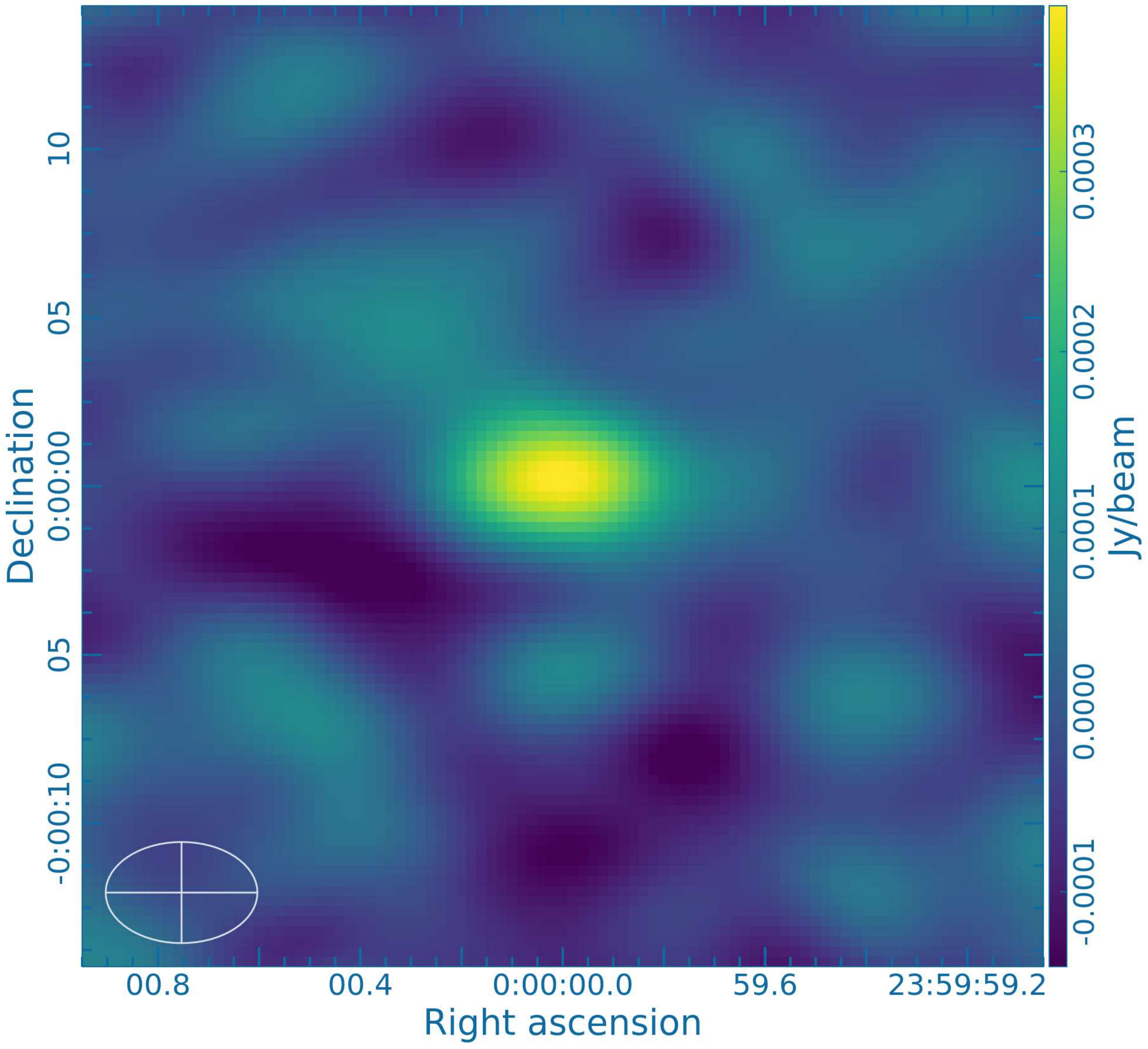}
    \caption{Stacked images of the continuum emission using two different convolution methods. \textbf{Left:} Continuum image stacked in the \emph{uv}-plane using the \texttt{ViSTa} method, then convolved with a largest beam of the sample. \textbf{Right:} Continuum image stacked in the image plane after convolving all input images to the worst resolution . The visibility-plane stacking yields a smoother background and captures a larger fraction of the source emission.}
    \label{fig:fig2_cont}
    
\end{figure}

We also tested the use of the \texttt{ViSTa} method to extract only the continuum emission and to quantify the potential gain provided by this approach. In this case, we produced the continuum image from the \texttt{ViSTa}-stacked measurement set, using only the channels identified as free from line emission. This approach was feasible because, in our simulated dataset, the line-free channels were known a priori. If applied to a realistic dataset, the continuum should instead be obtained by subtracting the line emission following the procedure described in Sect. \ref{sec:continuum}.

The continuum analysis was carried out following the procedure described in the preceding paragraph, homogenizing all data to the lowest angular resolution of the dataset. This ensures a realistic and coherent noise estimation for both methods, facilitating a fair comparison and enabling the quantification of the resulting improvement in SNR. While for the line stacking we generated the spectral cubes with \texttt{TCLEAN} using the \texttt{spemode='cube'} parameter, for the continuum we used  \texttt{spemode='mfs'}(\emph{multi-frequency synthesis}) and selected only the channels containing line-free continuum emission. The results are presented in Fig. \ref{fig:fig2_cont}. For both approaches, we measured the peak flux within a beam size region centered on the source. 
We note, however, that the peak flux does not represent the total flux density of the source, as the new restoring beam is comparable in size to the source itself and not significantly larger, as required for a point-like case. Nevertheless, it ensures that the SNR difference is driven solely by the gain of the two methods rather than by uncertainties in size estimation.
Using the \texttt{ImSta}–\texttt{ImFit} method, the peak flux is $0.404$ mJy/beam with an RMS noise of $0.058$ mJy/beam, giving a 
SNR of 6.966. With \texttt{ViSta}–\texttt{ImFit}, the peak flux is $0.419$ mJy/beam and the RMS noise $0.051$ mJy/beam, resulting in a SNR of 8.216. This represents a $\sim$18\% increase in peak SNR, consistent with the 1.2 gain reported in recent ALMA archival combination analysis which compares \emph{uv}-plane and image-plane stacking \citep[]{Hill2024}.

\section{Application to a real case}\label{sec:real_data}

After validating the method on the simulated case, we now apply the same analysis to a sample of galaxies observed with ALMA, targeting the CO(3–2) line emission. The sample consists of 21 partially resolved or unresolved galaxies, four of which had two available observations, in the redshift range of $z\sim 1.407 - 2.73$ and angular resolutions between 0.25 and 2.8 $^{\prime\prime}$. For each source, we verified the accuracy of its coordinates and ensured that every observation was free from instrumental issues or artifacts.

\begin{deluxetable*}{lcccccc}[!htpb]
\tabletypesize{\scriptsize}
\tablewidth{0pt}
\tablecaption{Source Properties \label{tab:source_properties}}
\tablehead{
\colhead{Source} & 
\colhead{$z_\mathrm{Spec}$} & 
\colhead{RA} & 
\colhead{Dec} & 
\colhead{$D_L$ factor} & 
\colhead{Ang Res ($^{\prime\prime}$)} &
\colhead{Peak SNR}
}
\startdata
HerBS-11    & 2.631     & 1:24:07.518   & -28:14:34.84  & 0.460     & 0.317     & 1.500 \\ 
HerBS-24    & 2.198     & 0:47:36.096   & -27:29:51.943 & 0.370     & 2.445     & 1.106 \\
HerBS-37    & 2.619     & 23:26:23.137  & -34:26:43.993 & 0.457     & 0.361     & \\
HerBS-45    & 2.434     & 0:51:32.943   & -30:18:49.566 & 0.418     & 2.366     & \\
HerBS-63    & 2.432     & 0:51:31.700   & -30:20:20.383 & 0.418     & 2.367     & 0.626 \\
HerBS-77    & 2.228     & 0:56:29.253   & -31:12:07.431 & 0.376     & 2.459     & 0.741 \\
HerBS-80    & 2.231     & 23:00:02.546  & -31:50:08.808 & 0.376     & 2.505     & \\
HerBS-86    & 2.564     & 23:53:24.571  & -33:11:11.766 & 0.446     & 2.661     & 1.008 \\
HerBS-93    & 2.400     & 23:47:50.447  & -35:29:29.932 & 0.411     & 0.250     & \\
HerBS-107   & 2.553     & 1:45:20.060   & -31:38:32.954 & 0.443     & 2.731     & \\
HerBS-135   & 2.401     & 22:56:11.799  & -32:56:51.802 & 0.412     & 0.348     & \\
HerBS-138   & 1.407     & 1:17:30.439   & -32:07:22.445 & 0.214     & 1.663     & 1.262 \\
HerBS-145   & 2.730     & 1:23:34.643   & -31:46:23.483 & 0.481     & 0.319     & \\
\textemdash & \textemdash & \textemdash & \textemdash   & \textemdash & 2.744   & \\
HerBS-159   & 2.236     & 23:51:21.740  & -33:29:00.400 & 0.377     & 2.669     & 1.756 \\
HerBS-178A  & 2.658     & 1:18:50.254   & -28:36:43.941 & 0.466     & 0.318     & \\
HerBS-178B  & 2.655     & 1:18:50.127   & -28:36:40.863 & 0.465     & 0.318     & \\
HerBS-182   & 2.227     & 23:05:38.804  & -31:22:05.351 & 0.376     & 2.471     & \\
HerBS-184   & 2.507     & 23:49:55.665  & -33:08:34.370 & 0.434     & 0.253     & \\
\textemdash & \textemdash & \textemdash & \textemdash   & \textemdash & 2.544   & 1.298 \\
HerBS-200   & 2.151     & 1:43:13.303   & -33:26:33.125 & 0.360     & 2.725     & \\
HerBS-208   & 2.478     & 22:57:44.560  & -32:42:33.566 & 0.428     & 1.527     & \\
\textemdash & \textemdash & \textemdash & \textemdash   & \textemdash & 0.350   & \\
HerBS-209   & 2.272     & 22:49:20.518  & -33:29:41.137 & 0.385     & 2.508     & \\
\textemdash & \textemdash & \textemdash & \textemdash   & \textemdash & 2.802   & \\ 
\enddata
\tablecomments{List of sources with redshift, coordinates, luminosity distance factor $D_L$, angular resolution of the observation and Peak SNR (in case of detection) included in the real sample described in \S \ref{sec:real_data}. When multiple observations exist for the same source, only the angular resolution is shown for secondary entries.}
\end{deluxetable*}

The selected sources are part of the sample analyzed in the BEARS I, II, and III studies \citep{bearsI,bearsII,bearsIII}, which provided spectroscopic redshifts through ALMA spectral scans, multi-frequency continuum observations to constrain the source spectral energy distributions (SEDs), and molecular line detections to characterize the ISM properties of these galaxies.
Detailed information for each source, including coordinates, resolution, and spectroscopic redshift, is provided in Table \ref{tab:source_properties}, along with the peak SNR value in the case of a spectral line detection. 

As we can estimate an approximate source size from the zero order moment maps for our sample, we used the same approach as the simulations (see Sect. \ref{sec:estimate_nativeres}). However, for the \texttt{ViSta}–\emph{uv}Fit case, we deliberately chose not to estimate the source size from the moment-zero map. We instead kept the spatial parameters of the Gaussian model free in each channel. This choice allows for greater flexibility, as line properties, such as velocity dispersion and extent, are not known a priori and may vary across the profile. Moreover, we did not subtract the continuum to avoid biasing the line emission. We applied a weighting scheme to account for varying frequency coverage across observations and to avoid multiple counting sources with duplicate observations, ensuring that each source contributes appropriately to the final stacked dataset. However, we did apply a correction for luminosity distance differences, adopting a reference redshift of 2.5 for consistent comparison with previous stacking analyses of high-$z$ DSFGs \citep{bearsIII, spilker2014}.

\begin{figure}[!htpb]
    \centering
    \includegraphics[trim={1.4cm 0.7cm 1.1cm 0.8cm},clip,width=0.98\linewidth]{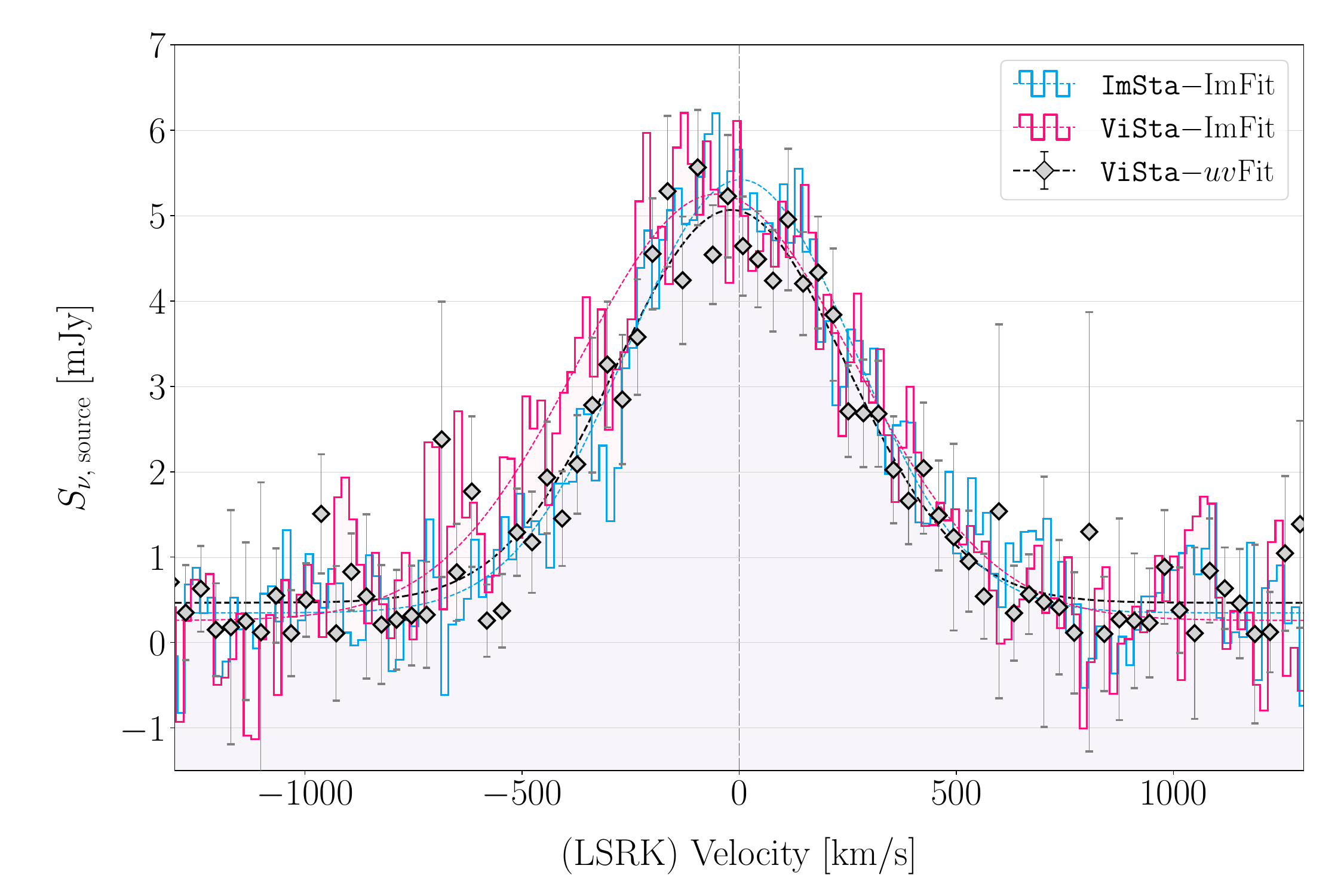}
    \caption{Stacked CO (3-2) spectra obtained through the different methods on the real sample data.  The black diamonds show the best-fit results from the \texttt{UVMultiFit} analysis with 1$\sigma$ uncertainties, and the blue line indicates the overall spectral fit to those data points. 
     The blue and the magenta lines show respectively the results from the \texttt{ImSta}-ImFit and \texttt{ViSta}-ImFit procedures, and the dashed lines represent the Gaussian fits to the data points.}
    \label{fig:CO32_fit}
\end{figure}

\begin{deluxetable}{lcccc}[!htpb]
\tabletypesize{\scriptsize}
\tablewidth{0pt}
\tablecaption{Gaussian fit parameters for the real sample}
\tablehead{
\colhead{Case} & 
 \colhead{$\mu$ (km/s)} & 
\colhead{$\sigma$ (km/s)} & 
\colhead{A (mJy)} & 
\colhead{C (mJy)} 
}
\startdata
\texttt{ViSta}-\emph{uv}Fit & -16.8184 & 264.9165 & 4.5835 & 0.4636 \\
\texttt{ViSta}-ImFit & -52.6187 & 317.8469 & 4.9921 & 0.2579 \\
\texttt{ImSta}-ImFit  & 3.3452 & 262.5233 & 5.0738 & 0.3467 \\
\enddata
\tablecomments{Summary of Gaussian fit parameters (mean, sigma, amplitude, shift) with uncertainties omitted for clarity. The values correspond to three different stacking methods applied to the real data sample.}
\label{tab:gaussian_fit_co32}
\end{deluxetable}

The results are presented in Figure \ref{fig:CO32_fit}. The various fitting methods appear to be reasonably consistent, as summarized in Table~\ref{tab:gaussian_fit_co32}. The velocity dispersion and the amplitude of the spectral line show good agreement across methods, with differences of approximately 10\% in both cases. The continuum level shows the largest relative variation, which is expected, given that the continuum is generally detected at a lower SNR than the line itself in most of our sample. In contrast, the position of the peak is consistent within 3 spectral channels across methods, which is a reasonable level of agreement considering the typical line widths and the possibility of minor asymmetries or extended emission that may affect the line differently. 
The resulting line properties are broadly consistent with other estimates in the literature \citep{bearsIII, spilker2014}. However, the result presented in this paper should not be interpreted as a precise measurement of the line emission, as we only applied a correction for the luminosity distance but did not account for other factors such as flux normalization or lensing magnification. This application to the real dataset is intended primarily as a test and comparison of the \texttt{ViSta} method with respect to both the simulated case and the classical stacking methods, rather than as a definitive estimate of the total line flux, which will be addressed in a future study.

\begin{figure}[!htpb]
    \centering
    \includegraphics[trim={0cm 0cm 0cm 0cm},clip,width=0.75\linewidth]{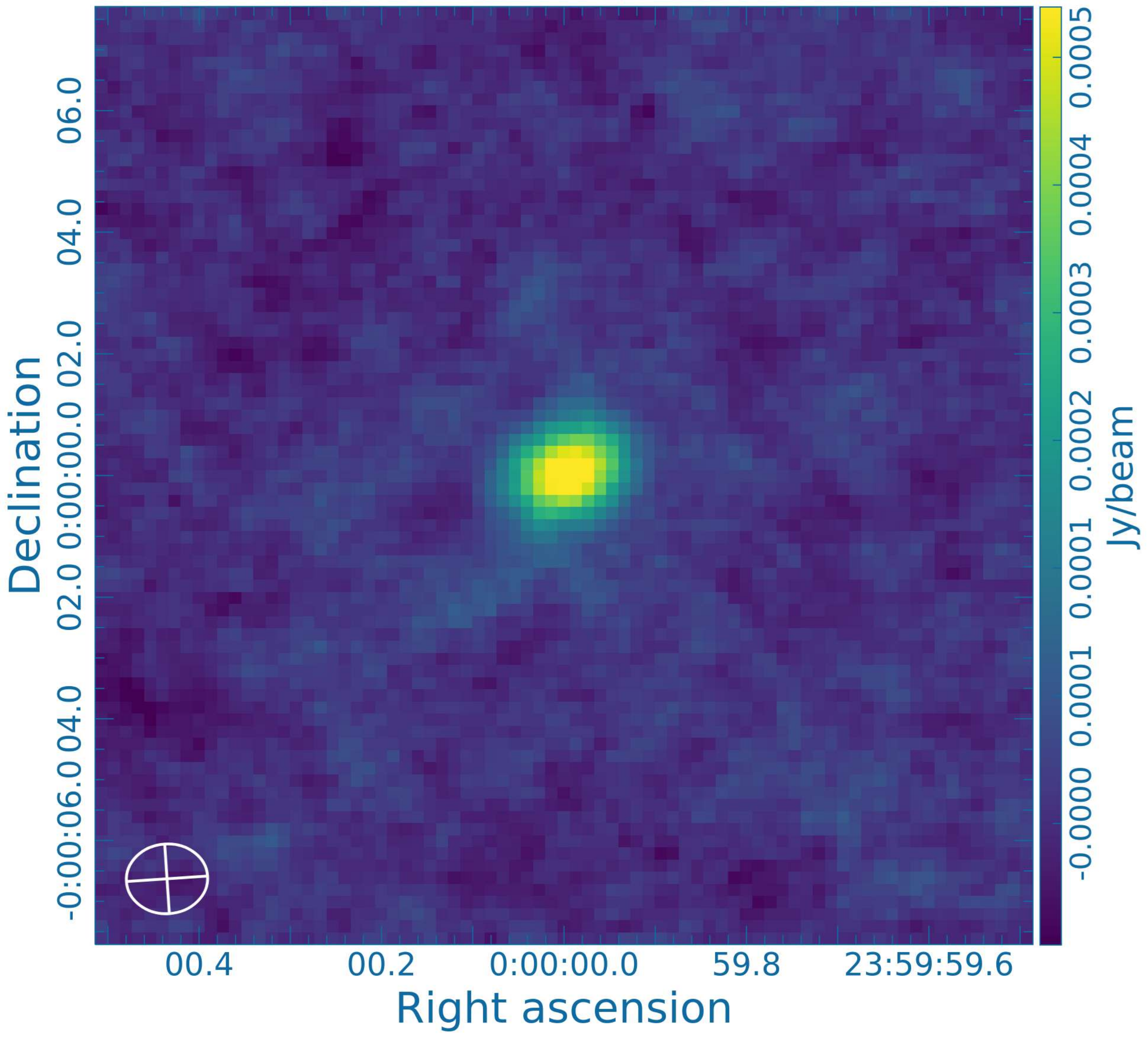}
    \caption{Stacked CO (3-2) continuum extracted as a post-processing step of the \texttt{ViSta} method after the concatenation of the visibilities. We estimate a continuum average flux density of $0.832 \pm 0.064$ mJy.}
    \label{fig:CO32_cont}
\end{figure}

As an additional test, we explored estimating the continuum as a post-processing step of the \texttt{ViSta} method. From the resulting continuum image, shown in Figure~\ref{fig:CO32_cont}, we estimate a continuum average flux density of $0.832 \pm 0.064$ mJy.

\begin{figure}[!htpb]
    \centering
    \includegraphics[trim={1.4cm 0.7cm 1.1cm 0.8cm},clip,width=\linewidth]{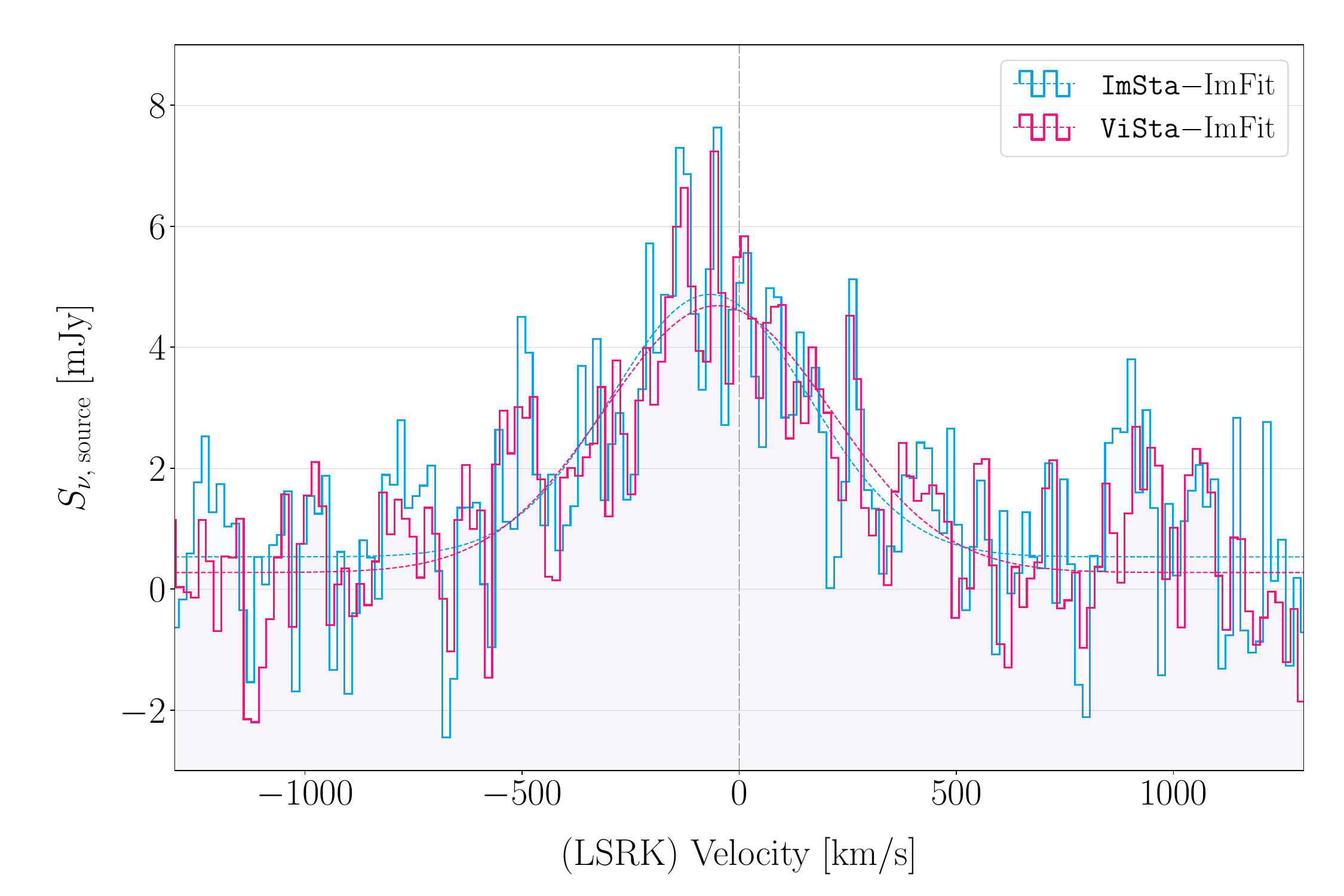}
    \caption{Stacked CO (3-2) spectra obtained through \texttt{ImSta}-ImFit and \texttt{ViSta}-ImFit procedures after convolving to the worst resolution of the dataset. The blue and the magenta lines show respectively the results from the \texttt{ImSta}-ImFit and \texttt{ViSta}-ImFit procedures, and the dashed lines represent the Gaussian fits to the data point.The spectrum derived from the \texttt{ImSta}-Imfit procedure appears noticeably more scattered and exhibits a higher level of noise fluctuations, whereas the \texttt{ViSta}-ImFit result shows a smoother and more coherent line profile, indicative of a more efficient suppression of random noise in the visibility-based stacking.}
    \label{fig:CO32_four}
\end{figure}

\begin{figure*}[!htpb]
    \centering
    \includegraphics[trim={1.1cm 0.7cm 3.8cm 0.8cm},clip,width=0.75\linewidth]{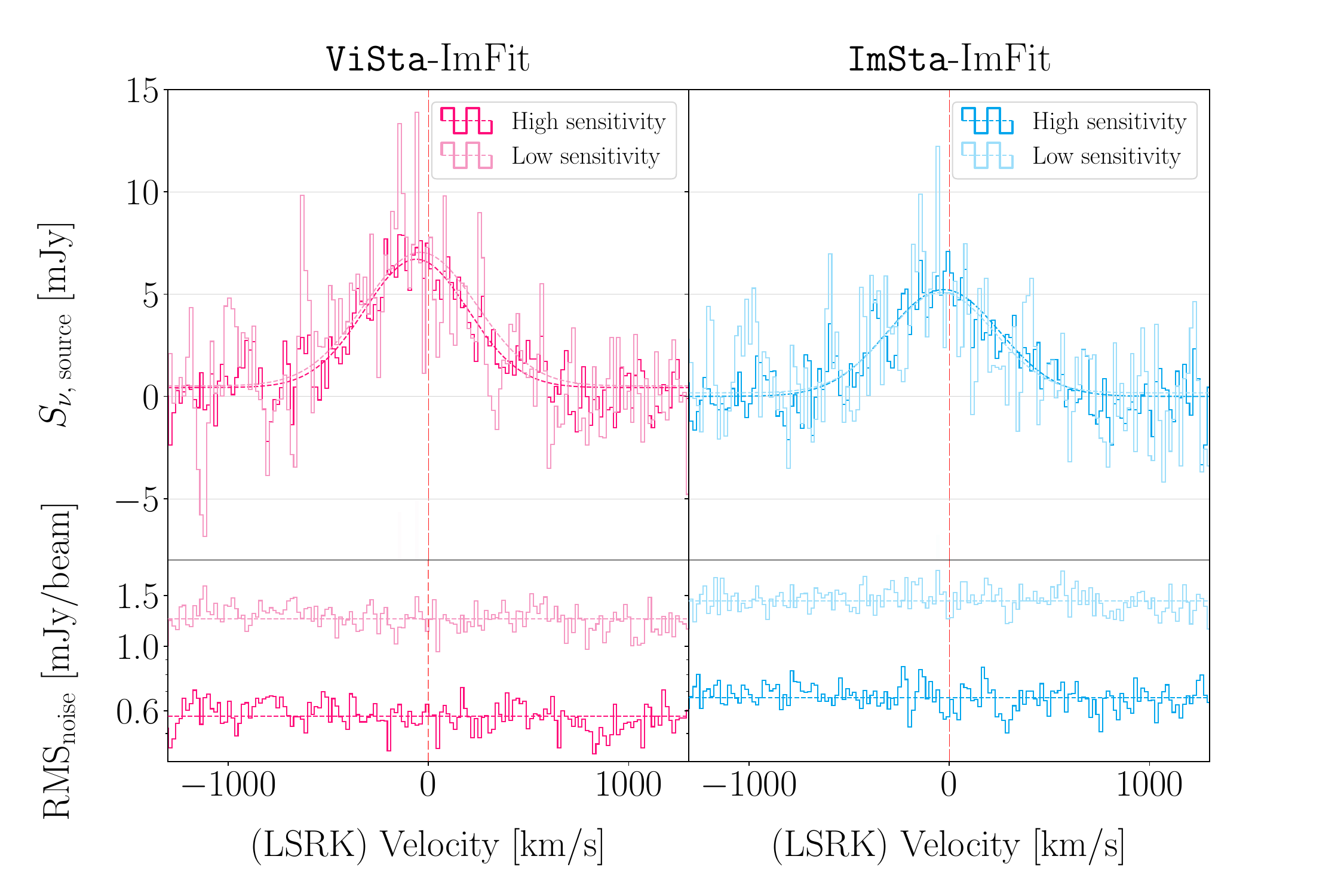}
    \caption{Comparison between stacking methods for a datasets with different sensitivities. \textbf{Top panels:} flux density extracted from the stacked data cubes at both high and low sensitivities. Both are shown with solid lines, with the low-sensitivity data in lighter colors, while the dashed lines represent the corresponding Gaussian fits. The left panel corresponds to the \texttt{ViSta}-ImFit, while the right panel shows the results from conventional \texttt{ImSta}-ImFit. \textbf{Bottom panels:} corresponding RMS noise as a function of velocity, derived from line-free channels in each cube. All datasets have been convolved to the same synthesized beam to ensure consistent noise estimation across methods. The spectra demonstrate that, at both sensitivity levels, \texttt{ViSta} yields a higher peak flux density and reduced noise fluctuations relative to \texttt{ImSta}, leading to a systematically higher SNR. The improvement becomes particularly evident in the low-sensitivity case, where the visibility-based stacking more effectively preserves the coherence of the faint line signal.}
    \label{fig:CO32_conv_compare}
\end{figure*}

In the case presented, all sources were detected from the first integration, and for nearly all targets the spectral line was clearly detected. This allowed for a robust characterization of the line parameters, which naturally explains why both stacking approaches produce consistent results.

A more stringent observational test of the capability of the \texttt{ViSta} method can be performed by artificially degrading the sensitivity of individual observations. This can be achieved by reducing the number of integration intervals, and thus the $uv$-coverage, which lowers the overall data quality. When a source is no longer individually detectable, we expect the \texttt{ViSta} method to recover a larger fraction of the intrinsic signal and achieve a higher SNR compared to conventional image stacking, as we demonstrated in the simulated case. To explore this, we selected one fourth of the available data for each observation and repeated the analysis, thereby testing the method under conditions of reduced sensitivity and sparser $uv$-sampling. 
In the first phase, we compared the recovered line flux between the classical stacking method and \texttt{ViSta} at native resolution. To determine the area over which to measure the flux density, we referred to the same moment-zero maps generated from the higher-sensitivity observations, as done in the previous case. The results, shown in Fig. \ref{fig:CO32_four}, indicate that when the sensitivity of the observations is reduced and the line signal is increasingly embedded in the noise, stacking in the image plane produces a source that appears significantly more scattered and diffuse compared to one in the \emph{uv}-plane. Thus, under conditions approaching the detection limit, \texttt{ViSta} seems to be more effective in capturing the line emission. 

However, to obtain a consistent estimate of the actual SNR improvement achieved by stacking in the visibility domain, it is necessary to evaluate the noise coherently for both the visibility and image-plane stacking. To do so, we refer to the procedure described in Sect. \ref{sec:lowest_res} and convolve all datasets to the worst angular resolution available, corresponding to a synthesized beam of $5.063^{\prime\prime} \times 2.820^{\prime\prime}$. Figure \ref{fig:CO32_conv_compare} illustrates the obtained results: the top panels show the spectra extracted using the \texttt{ViSta}-ImFit and \texttt{ImSta}-ImFit methods for both the high- and low-sensitivity datasets, while the bottom panels display the corresponding RMS noise.

When considering the datasets at their original sensitivity, stacking in the \emph{uv}-plane yields a more effective suppression of noise, consistent with the behaviour observed in the simulated tests. Quantitatively, the SNR increases from 7.837 for \texttt{ImSta}-ImFit to 10.924 for \texttt{ViSta}-ImFit, corresponding to an improvement of approximately 39\%. When the data sensitivity is reduced, the SNR increases from 3.41 to 5.24, resulting in a relative gain of about 53\%. These results confirm our earlier hypothesis that visibility-based stacking becomes progressively more advantageous as the data approach the noise-dominated regime, owing to its ability to preserve the phase coherence of the astronomical signal while mitigating the incoherent addition of noise inherent to image-plane stacking.

\section{Summary and outlook}\label{sec:conclusions}

In this paper, we have introduced \texttt{ViSta}, a novel technique for stacking radio-to-submm data directly in the \emph{uv}-domain. This method enables the combination of observations of galaxies at different redshifts, obtained with various telescopes and array configurations, into a unified analysis framework. 

We have outlined the advantages of performing stacking in the visibility plane according to the \texttt{ViSta} method, presented the related workflow pipeline, and demonstrated its efficiency on simulated and real ALMA observations for the study of the CO(3-2) average line emission on a sample of dusty star-forming galaxies.

The main results of our analysis are listed as follows:
\begin{itemize}
    \item Overall, the \texttt{ViSTa} method performs correctly and reliably when tested with the simulated dataset.  While for point-like sources its performance is comparable to classical image-plane stacking, the true advantage becomes evident in the case of extended sources, where the intrinsic flux and size cannot be reliably recovered in individual images because of the low SNR. In such cases, performing the fit directly in the visibility domain after stacking yields a more accurate estimate of the total flux and a lower noise level compared to the approach of fitting each individual image and then averaging the recovered flux densities. When applied to extract only the continuum emission, the \texttt{ViSTa} method provides a gain of about 18\% over classical image-plane stacking in our analysis. This highlights the strength of visibility-domain stacking in preserving faint, spatially extended emission that would otherwise be underestimated.

    \item While testing \texttt{ViSTa} with real data of high-$z$ DSFGs observed with ALMA, where the spectral line was detectable in some sources or where a reliable zero-moment map enabled fitting of source sizes, both the \texttt{ViSTa} visibility-plane fitting and classical image-plane stacking methods yield similar flux estimates. This consistency confirms the validity of our approach under conditions where the source morphology can be reliably constrained, and it is in agreement with simulations and previous stacking experiments on local and Galactic sources \citep{hancock2011, neumann2023}. The fluxes derived from image-plane stacking also agree with prior literature values, further validating our methodology. However, we note that the physically correct total flux would still require additional normalization steps, such as proper flux calibration and correction for lensing magnification or selection biases, which are beyond the scope of this paper.
     \item To further test our method, we artificially reduced the sensitivity of the observations by using only one fourth of the data, lowering $uv$-coverage and overall data quality. Under these conditions, where individual sources are no longer easily detectable, \texttt{ViSTa} recovers a larger fraction of the intrinsic signal and achieves higher SNR compared to image-plane stacking. At native resolution, image-plane stacking produces more scattered and diffuse sources, while \texttt{ViSTa} preserves the line emission more effectively. When all datasets are convolved to the worst angular resolution, this advantage remains evident, with \texttt{ViSTa} achieving higher SNR than image-plane stacking increasing from a relative gain of approximately 39\% in the high-sensitivity case to about 53\% in the reduced-sensitivity case, confirming the method’s robustness under low-SNR and sparse $uv$-sampling regimes.

\end{itemize}

In the future, the \texttt{ViSTa} method will enable the reconstruction of average molecular line properties, such as SLEDs and line ratios, by stacking ALMA archival data of DSFGs, offering a unique window on the bulk physical conditions of the molecular gas in these systems. For instance, the relative intensities of different CO rotational lines, among the brightest spectral features in these sources, reveal the CO level populations, which depend sensitively on gas temperature and density; this can lead to the inference of the total molecular gas mass under the derived excitation, showing whether these extreme starbursts exhibit nearly thermalized CO ladders (indicative of very warm, dense gas) or contain large amounts of cooler, sub-thermally-excited gas – hence refining our estimates of the ISM conditions and gas content (Torsello et al. in prep.).

By analyzing the low and high excitation level ratios it will be possible to investigate the relative role of nuclear activity (responsible for the highest density and temperature) and star formation. By matching them with other spectral features one could try to select sub-samples and investigate how such relations evolve within the diverse populations across  times. 

Beyond the brighter emission lines, the power of visibility domain stacking will enable searches for much fainter lines. For example, one could target water (H$_2$O) lines, which can dominate the cooling of hot, dense molecular gas and intense, dust-enshrouded star formation. We will also employ \texttt{ViSta} to investigate shock tracers (such as HCN or CS) or to search in high-redshift galaxies for intrinsically faint lines that have so far only been detected in the local Universe (i.e. methanol, a tracer of grain-surface chemistry and turbulence).
A public release of the basic \texttt{ViSta} code tailored to the ASA will be made available at the time of publication. The code repository is hosted at \href{https://github.com/martitors/ViSta}{\texttt{https://github.com/martitors/ViSta}}.

Furthermore, an advanced version is planned, where particular attention will be given to minimizing the use of computationally expensive operations and integrating high-performance computing (HPC) strategies to handle large datasets efficiently. This is especially crucial given that stacked data products can exceed several hundred gigabytes, far beyond the handling capabilities of standard pipelines or tools designed for individual telescope datasets. In the advanced version of the code, the program will aim to rely on CASA only for calibration, while data extraction and all subsequent operations described in the pipeline will be implemented in a more efficient and optimized way. In this manner, once the \emph{uv}-plane data are extracted, the workflow becomes identical for datasets from any interferometric telescope, such as the VLA, NOEMA, ATCA, ALMA, and MeerKAT. Even if the raw data are initially stored in different formats, they can be extracted using their native software packages and subsequently combined within the same unified framework, enabling the simultaneous stacking of observations from multiple facilities. Looking ahead, tools like ViSta will be particularly valuable in the SKA era, where the sheer volume and complexity of interferometric data will demand scalable and efficient analysis solutions.

\begin{acknowledgments}
This work was partially funded from the projects: INAF GO-GTO Normal 2023 funding scheme with the project "Serendipitous H-ATLAS-fields Observations of Radio Extragalactic Sources (SHORES)"; INAF Large Grant 2022 project "MeerKAT and LOFAR Team up: a Unique Radio Window on Galaxy/AGN co-Evolution; INAF Large GO 2024 project "MeerKAT and Euclid Team up: Exploring the galaxy-halo connection at cosmic noon"; ``Data Science methods for MultiMessenger Astrophysics \& Multi-Survey Cosmology'' funded by the Italian Ministry of University and Research, Programmazione triennale 2021/2023 (DM n.2503 dd. 9 December 2019), Programma Congiunto Scuole; EU H2020-MSCA-ITN-2019 n. 860744 \textit{BiD4BESt: Big Data applications for black hole Evolution STudies}; Italian Research Center on High Performance Computing Big Data and Quantum Computing (ICSC), project funded by European Union - NextGenerationEU - and National Recovery and Resilience Plan (NRRP) - Mission 4 Component 2 within the activities of Spoke 3 (Astrophysics and Cosmos Observations); ORP, that is funded by the European Union’s Horizon 2020 research and innovation programme under grant agreement No 101004719 [ORP]. This paper makes use of the following ALMA data: \texttt{ADS/JAO.ALMA\#2019.1.01477.S} (P.I.: S. Urquhart), \texttt{ADS/JAO.ALMA\#2021.1.01628.S} (P.I.: T. Bakx), and \\
\texttt{ADS/JAO.ALMA\#2022.1.00432.S} (P.I.: T. Bakx). ALMA is a partnership of ESO (representing its member states), NSF (USA) and NINS (Japan), together with NRC (Canada), NSTC and ASIAA (Taiwan), and KASI (Republic of Korea), in cooperation with the Republic of Chile. The Joint ALMA Observatory is operated by ESO, AUI/NRAO and NAOJ. 
\end{acknowledgments}

\begin{contribution}
All authors contributed equally to the planning of the \texttt{ViSta} method, to the discussion of the results and to the writing of the present manuscript. 
MT is responsible of the development of the \texttt{ViSta} code, and performed the described simulations and analysis.s.

\end{contribution}

\facilities{ALMA}

\software{\texttt{Astropy} \citep{astropy},  
          \texttt{Scipy} \citep{Scipy},
          \texttt{Matplotlib} \citep{matplotlib},
          \texttt{CASA} \citep{casa},
          \texttt{CARTA} \citep{carta},
          \texttt{UVMultiFit} \citep{uvmultifit}
          }

\bibliography{bibliography}{}
\bibliographystyle{aasjournalv7}

\end{document}